\documentclass[]{aastex631}

\newcommand\aastex{AAS\TeX}

\usepackage{mathtools}
\usepackage{graphicx}
\usepackage{subcaption}
\usepackage{ragged2e}
\DeclareCaptionJustification{justified}{\justifying}
\usepackage[labelfont=bf]{caption} % for bold figure captions in sidecap
\usepackage{sidecap} % for captions next to plot
\usepackage{wrapfig} % for text that wraps around tables and figures

\begin{document}

\title{Template \aastex Article with Examples: 
v6.3.1\footnote{Released on March, 1st, 2021}}

\title{Evolution of Relativistic Pair Beams: Implications for Laboratory and TeV Astrophysics}

\author{Marvin Beck}
\affiliation{II. Institut f\"ur Theoretische Physik, Universit\"at Hamburg, 
Luruper Chaussee 149, 22761 Hamburg, Germany}

\correspondingauthor{Oindrila Ghosh}
\email{oindrila.ghosh@fysik.su.se \\ oindrila.ghosh@desy.de}

\author[0000-0003-2226-0025]{Oindrila Ghosh}
\affiliation{Stockholm University and The Oskar Klein Centre for Cosmoparticle Physics, Alba Nova, 10691 Stockholm, Sweden}
\affiliation{II. Institut f\"ur Theoretische Physik, Universit\"at Hamburg, 
Luruper Chaussee 149, 22761 Hamburg, Germany}

\author{Florian Gr\"uner}
\affiliation{Institut f\"ur Experimentalphysik, Universit\"at Hamburg, Luruper Chaussee 149, 22761 Hamburg, Germany}
\affiliation{Center for Free-Electron Laser Science, Notkestraße 85, DE-22607 Hamburg, Germany}

\author{Martin Pohl}
\affiliation{Institute of Physics and Astronomy, University of Potsdam, D-14476 Potsdam, Germany}
\affiliation{Deutsches Elektronen-Synchrotron DESY, Platanenallee 6, 15738 Zeuthen, Germany}

\author{Carl B. Schroeder}
\affiliation{Lawrence Berkeley National Laboratory, 1 Cyclotron Rd, Berkeley, California 94720, USA}
\affiliation{Department of Nuclear Engineering, University of California, Berkeley, California 94720, USA}

%Center for Free-Electron Laser Science, DESY, Notkestraße 85, DE-22607 Hamburg, Germany 

\author[0000-0002-4396-645X]{G\"unter Sigl}
\affiliation{II. Institut f\"ur Theoretische Physik, Universit\"at Hamburg, Luruper Chaussee 149, 22761 Hamburg, Germany}

\author{Ryan D. Stark}
\affiliation{Institut f\"ur Experimentalphysik, Universit\"at Hamburg, Luruper Chaussee 149, 22761 Hamburg, Germany}
\affiliation{Center for Free-Electron Laser Science, Notkestraße 85, DE-22607 Hamburg, Germany}

\author{Benno Zeitler}
%\affiliation{Deutsches Elektronen-Synchrotron DESY, Notkestraße 85, 22607 Hamburg, Germany}
\affiliation{Institut f\"ur Experimentalphysik, Universit\"at Hamburg, Luruper Chaussee 149, 22761 Hamburg, Germany}
\affiliation{Center for Free-Electron Laser Science, Notkestraße 85, DE-22607 Hamburg, Germany}
% Ask Florian and Ryan if they also want the CFEL Affilation

%Center for Free-Electron Laser Science, DESY, Notkestraße 85, DE-22607 Hamburg, Germany 

\begin{abstract}

Missing cascades from TeV blazar beams indicate that collective plasma effects may play a significant role in their energy loss. It is possible to mimic the evolution of such highly energetic pair beams in laboratory experiments using modern accelerators. The fate of the beam is governed by two different processes, energy loss through the unstable mode and energetic broadening of the pair beam through diffusion in momentum space. We chalk out this evolution using a Fokker-Planck approach in which the drift and the diffusion terms respectively describe these phenomena in a compact form. We present particle-in-cell simulations to trace the complete evolution of the unstable beam-plasma system for a generic narrow Gaussian pair beam for which the growth rate is reactive. We show that the instability leads to an energetic broadening of the pair beam, slowing down the instability growth in the linear phase, in line with the analytical and numerical solutions of the Fokker-Planck equation. Whereas in a laboratory experiment the change in the momentum distribution is an easily measured observable as a feedback of the instability, the consequence of diffusive broadening in an astrophysical scenario can be translated to an increase in the opening angle of the pair beam. 

\end{abstract}

\keywords{TeV blazars (x) --- Plasma instabilities (x) --- Laboratory astrophysics (x) --- PIC simulation(x)}

%\begin{document}
\section{Introduction} \label{sec:intro}

Among the known extragalactic gamma-ray sources, blazars are ubiquitous and numerous in the TeV sky. Blazars are typically active galactic nuclei with their jets pointing towards the line of sight and TeV emissions from such distant sources are attenuated through their interaction with the extragalactic background light (EBL), leading to the generation of electron-positron pair beams propagating in the same direction as the emitted TeV photons. Subsequent energy losses of the pair beam are typically attributed to inverse Compton scattering (ICS) when the beam interacts with the cosmic microwave background (CMB) producing GeV gamma rays. However, a discrepancy between the expected \citep{aharonian2001tev} and observed flux through various gamma-ray observations of blazars have been reported \citep{ neronov2010evidence, taylor2011}. This discrepancy, dubbed as the GeV-TeV tension, is widely known and is understood in terms of collective plasma effects \citep{broderick2012cosmological, broderick2014implications, miniati2013relaxation, supsar2014plasma, schlickeiser2013plasma, sironi2014relativistic, alawashra2022suppression,  kempf2016, rafighi2017plasma, vafin2018electrostatic, vafin2019, alves2019, shalaby2020} or deflection of the electron-positron pairs by the intergalactic magnetic field (IGMF). The non-observation of the predicted gamma-ray flux can be used to derive a lower bound on the strength of the IGMF, provided plasma instabilities do not play a significant role \citep{neronov2009sensitivity, neronov2010evidence, lemoine2010electromagnetic}. In addition, electromagnetic cascades in voids can, under the influence of the IGMF, produce extended emission, also known as ``pair halos'', i.e., bow-tie-like structures around point sources in the gamma-ray sky \citep{broderick2016bow}, although this has not been observed. Recent Fermi-LAT measurements of the isotropic gamma-ray background indicates that in the case that the IGMF is feeble, considering collective plasma effects are necessary to alleviate the tension \citep{Blanco:2023kfa}.
 
 Cherenkov resonance arising from the interaction between the pair beam and the background plasma leads to the growth of unstable modes. Such instabilities can be electrostatic or electromagnetic in nature \citep{bret2005electromagnetic}. The growth rate of such instabilities, and thus, the fate of the pair beam depends on the initial distribution function of particles in the pair beam. In this context, we examine the role of plasma instabilities in how pair beams can lose energy during propagation in comparison to ICS. It is, however, unclear whether the instabilities can contribute to significant energy drain \citep{sironi2014relativistic}, which has observable cosmological implications. For example, the injection of energy into the IGM plasma can affect the thermal history and suppress the formation of structures at small scales \citep{chang2012cosmological, broderick2012cosmological, Pfrommer:2013zla}. However, the role of plasma instabilities as an energy loss mechanism for the pair beam is under debate as \citep{perry2021role} and  \citep{alawashra2022suppression} respectively argued that inhomogeneities in the IGM plasma and weak tangled intergalactic magnetic field can render the instabilities inefficient for astrophysical pair beams.
 
 In this article, we consider a laboratory-based setup of a beam-plasma system and examine its fate through analytical estimates and particle-in-cell (PIC) simulations. In exploring the role of collective plasma processes, it is crucial to understand the temporal evolution of the beam-plasma system and how it varies while subject to changes in the key parameters. We find that the pair beam in question goes through three stages of evolution: i. the linear growth phase in which the unstable modes transfer energy from the beam into the background plasma, a dissipative process, ii. the diffusive relaxation phase, in which the beam undergoes broadening in the momentum space, and finally, iii. the nonlinear phase characterised by Landau damping of the modes, eventually leading to a saturation of instabilities. 

 Since the extent of the energy injection into the plasma is dependent on the efficiency of the plasma instabilities, in the first part of the article, we focus on identifying the oscillation modes that are important in driving the instability, sensitivity of the growth rate with respect to the relevant parameters, and how the energy densities in the beam and the plasma evolve with the instability growth. Later on, we demonstrate that the neutral pair beam not only loses energy to the plasma but also heats itself, characterised by an energetic widening of the beam in directions parallel and perpendicular to its propagation. For the first time, we present the dissipative and diffusive processes of plasma heating and beam self-heating respectively in a closed-form Fokker-Planck equation and present analytical and numerical solutions, respectively for 1D and 2D Gaussian input beams, and compare them with the PIC simulation results for similar parameters.
 
 In Section \ref{sec:plasmainst}, we outline the linear growth of beam-plasma instabilities, then in Section \ref{sec:diffusion} we describe the quasilinear relaxation, and finally in Section \ref{sec:pic} we report the findings from a particle-in-cell (PIC) simulation for a beam-plasma system achievable in a laboratory setting. Next, in Section \ref{sec:discussions}, we compare the analytical predictions, numerical solution of the Fokker-Planck equation, and results from the PIC simulation. To conclude,  we lay out the scope of our investigation and key takeaways in Section \ref{sec:conclusion}.
 
 \section{Plasma Instabilities and Growth of Modes} \label{sec:plasmainst}

From the interaction between the beam and the plasma, a variety of instabilities can develop based on beam parameters and distribution functions. % relevant for astrophysical, laboratory-based and fusion setups. 
For an arbitrary beam velocity characterized by $\mathbf{v}$, and wave vector $\mathbf{k}$, the two-stream mode ($\mathbf{k}\cdot\hat{\mathbf{v}}= \omega$) , the filamentation mode ($\mathbf{k}\cdot\hat{\mathbf{v}}=0$), and the general oblique mode ($\mathbf{k}\cdot\hat{\mathrm{v}}=\omega \cos\theta$) have been explored \citep{bret2008exact, bret2010exact,Bret:2010}. In addition to (semi-)analytical estimates of growth rates, PIC simulations initialized with arbitrary electron distributions have been useful in understanding the evolution of the beam, with the following parameters of significance: the density contrast between the beam and the plasma, the Lorentz factor of the beam, and the energy or momentum spread of the beam \citep{sironi2014relativistic, rafighi2017plasma}.

For unstable modes, the associated Langmuir waves satisfy the resonance condition:
 
 \begin{equation}
 \omega-\mathbf{k} \cdot \mathbf{v}=0.
 \end{equation}
 
 where $\omega=\omega(\mathbf{k})$ is the complex frequency. The imaginary part of the frequency $\mathrm{Im}\left(\omega (\mathbf{k})\right)$ signifies the instability growth rate. In order to determine the growth rate, one can solve the dispersion relation for a given distribution function, applying a linearization scheme. For a normalized phase-space distribution function:
 
 \begin{equation}
    F(\mathbf{p},\mathbf{x})/n_\text{b} = f(\mathbf{p}) =  f(p_{x}) f(p_{y}) f(p_{z}),
\end{equation}

where 

\begin{equation} 
    \int d^3 p f(\mathbf{p}) =
    \int_{-\infty}^{+\infty}dp_{x} f(p_{x})   \int_{-\infty}^{+\infty}dp_{y} f(p_{y}) \int_{-\infty}^{+\infty}dp_{z} f(p_z) = 1,
\label{eq:dfcart}
\end{equation}

and

\begin{equation}
    \mathbf{k}=(k_x, k_y, k_z),
\end{equation}

such that the beam distribution function is expressed in terms of separable components in each direction. For the moment we ignore any cross-terms that may be related to beam emittance.

Since axisymmetry may be assumed in an electron-positron pair beam without any loss of generality, the wave number tensor can now be represented in terms of a component parallel to the direction of propagation of the beam, $k_{\mid \mid} = k_z $, and a component perpendicular to the beam propagation, $k_{\perp}=(k_x^2 + k_y^2)^{1/2}$ \citep{schlickeiser2013plasma}.

Eq.~\ref{eq:dfcart} can then be recast as 

\begin{equation}
    \int_{-\infty}^{+\infty}d^3 p f(\mathbf{p})= \int_{0}^{+\infty}dp_{\perp} f(p_{\perp}) p_{\perp}  \int_0^{2\pi} d\phi \int_{-\infty}^{+\infty}dp_{\mid \mid} f(p_{\mid \mid}) = 1.
\end{equation}

In order to correctly describe the collective plasma phenomena, it is important to distinguish between various canonical scenarios: i.\ cold (monochromatic) beam propagating through cold plasma, ii.\ cold beam propagating through warm plasma, and iii.\ a warm beam with finite momentum spread propagating through warm background plasma. 
For a monoenergetic cold beam propagating in a background plasma whose constituents can be approximated to have vanishing velocity, the dispersion relation reduces to its hydrodynamic form, which can also be derived from the continuity and momentum equations,

\begin{equation}
1-\frac{\omega_\text{p}^{2}}{\omega^{2}}-\frac{\omega_\text{b}^{2}}{\gamma^{3}\left(\omega-k_{\mid \mid} v_\text{b}\right)^{2}} \frac{\gamma^{2} k_{\perp}^{2}+k_{\mid \mid}^{2}}{k_{\perp}^{2}+k_{\mid \mid}^{2}}=0,
\end{equation}

where the subscript p stands for the background plasma, and b represents the beam, $\omega_\text{p}$ and $\omega_\text{b}$ denote the plasma frequencies corresponding to the electron densities $n_\text{p}$ and $n_\text{b}$, respectively, and $\gamma$ denotes the Lorentz factor of the beam particles in the background plasma rest frame. This is also known as the reactive regime, characterised by the condition

 \begin{equation}
 \left| \mathbf{k} \cdot \Delta \mathbf{v} \right| \ll \left|\mathrm{Im} \left( \omega(\mathbf{k}) \right)\right|,
 \label{eq:kinvsreac} 
 \end{equation}
which for relativistic beams reduces to
 
\begin{equation}
\frac{\Delta p_{\mid\mid} }{ p}\ll \gamma^2\frac{\left|\mathrm{Im} \left(\omega(\mathbf{k})\right)\right|}{\omega_{p}}\quad\textrm{and}\quad 
\frac{\Delta p_{\perp} }{ p}\ll \frac{\left|\mathrm{Im} \left(\omega(\mathbf{k})\right)\right|}{\left(\frac{k_\perp c}{\omega_{p}}\right)\omega_{p}}
\end{equation}
  for the momentum components $p_{\mid\mid}$ and $p_\perp$ respectively parallel and perpendicular to $\mathbf{k}$.   
 Thus, when the momentum spread of the beam is sufficiently small, the instabilities occur in the reactive regime, otherwise they develop in the kinetic regime. The maximum reactive growth rate can be obtained as $\mathrm{Im}(\omega(k))= \delta_{r} \omega_{p}$ such that \citep{fainberg1972nonlinear}, 

\begin{equation}
\delta_{\mathrm{r}}=\frac{\sqrt{3}}{2^{4 / 3}}\left(\frac{2 \alpha}{\gamma}\right)^{1 / 3}\left(\frac{k_{\|}^{2}}{\gamma^{2} k^{2}} +\frac{k_{\perp}^{2}}{k^{2}}\right)^{1 / 3} = \frac{\sqrt{3}}{2^{4 / 3}}\frac{(2 \alpha)^{1/3}}{\gamma}(\cos{\theta_{0}^2}+\gamma^2 \sin{\theta_{0}}^2)^{1/3},
\label{eq:grreac}
\end{equation}

where $\theta_{0}$ is the initial angle between the wave vector and beam velocity, $\alpha = n_\text{b}/n_\text{p}$ is the density contrast, with $n_\text{b}$ and $n_\text{p}$ representing the number density of particles in the pair beam and the background plasma, respectively. In the reactive regime, the maximum growth rate occurs for the perpendicular modes. Furthermore, the development of instabilities in the kinetic regime can be examined for realistic beam distributions, such as an astrophysical pair beam and even a monochromatic momentum distribution that undergoes energetic broadening during the course of its evolution.

To obtain the instability growth rate in the oblique mode for warm beams with finite momentum spreads propagating through a background warm plasma, Eq.~\ref{eq:kinvsreac} does not hold and the corresponding calculation must be performed in the kinetic regime. For ultrarelativistic beams ($\upsilon \sim c$) with arbitrary beam distribution function cast in polar coordinates, $f(p, \theta)$, the normalised growth rate reduces to \citep{brejzman1974powerful},

\begin{equation}
\delta_{k}=- \pi \frac{n_\text{b}}{n_\text{p}}\left(\frac{\omega_\text{p}}{k c}\right)^{3} \int_{\mu_{-}}^{\mu_{+}} d \mu \frac{2 g+\left(\mu-\frac{k_{\|} c}{\omega_\text{p}}\right) \frac{\partial g}{\partial \mu}}{\left[\left(\mu_{+}-\mu\right)\left(\mu-\mu_{-}\right)\right]^{\frac{1}{2}}},
\label{eq:grkinetic}
\end{equation}

where

\begin{equation}
g(\theta)=\frac{m_{e} c}{n_\text{b}} \int dp p f(p, \theta),
\end{equation}

\begin{equation}
\mu_{\pm}= \frac{\omega_\text{p}}{k c} \left(\frac{k_{\|}}{k} \pm \frac{k_\perp}{k} \sqrt{\frac{k^{2} c^{2}}{\omega_\text{p}^{2}}-1}\right).
\end{equation}

In a possible laboratory experiment, the initially injected beam distribution function can be described by a 2D Gaussian distribution given by 

\begin{equation}
f(p_{\|}, p_{\perp})=\frac{1}{2 \pi \sigma_{\|} \sigma_{\perp}} \exp \left[-\frac{1}{2}\left(\frac{p-\mu_{\|}}{\sigma_{\|}}\right)^2-\frac{1}{2}\left(\frac{p_{\perp}-\mu_{\perp}}{\sigma_{\perp}}\right)^2\right].
\label{eq:2Dnormal}
\end{equation}

\begin{figure}[htp]

\centering
\includegraphics[width=.3\linewidth]{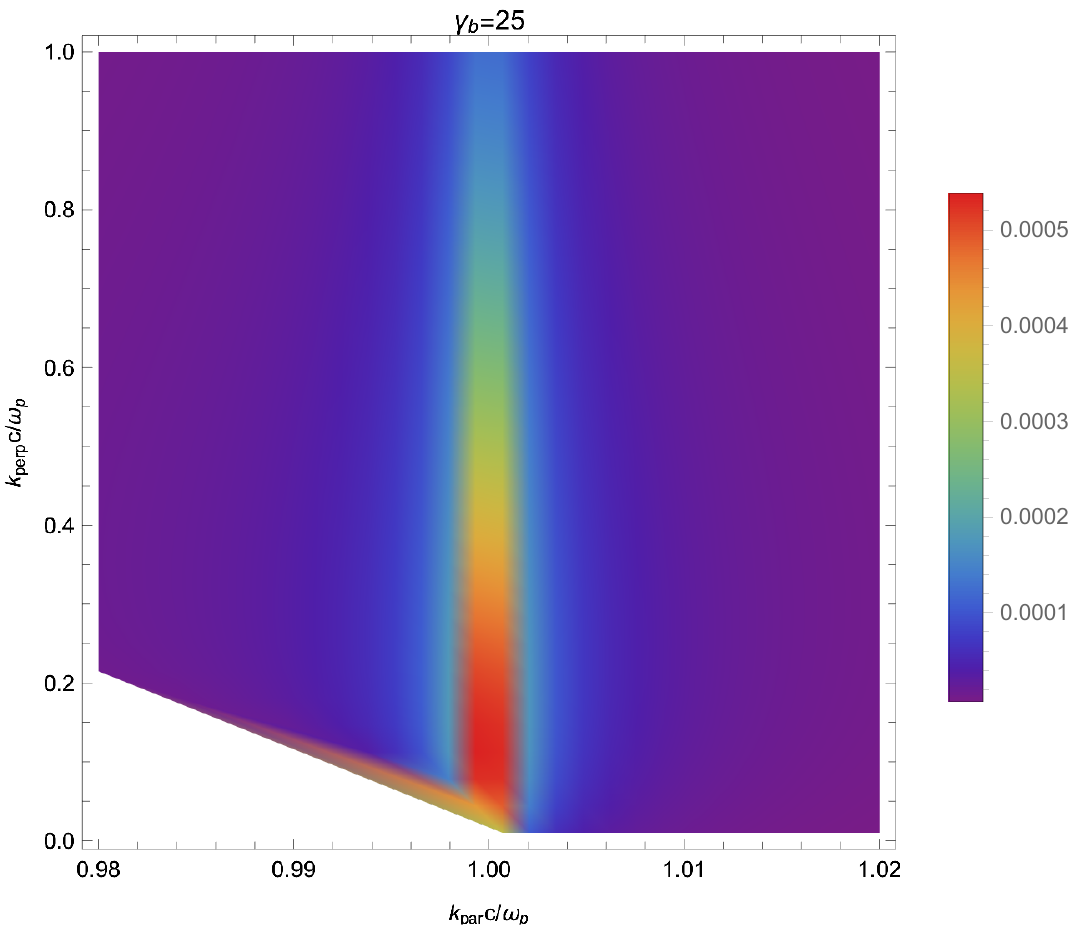}
\includegraphics[width=.3\linewidth]{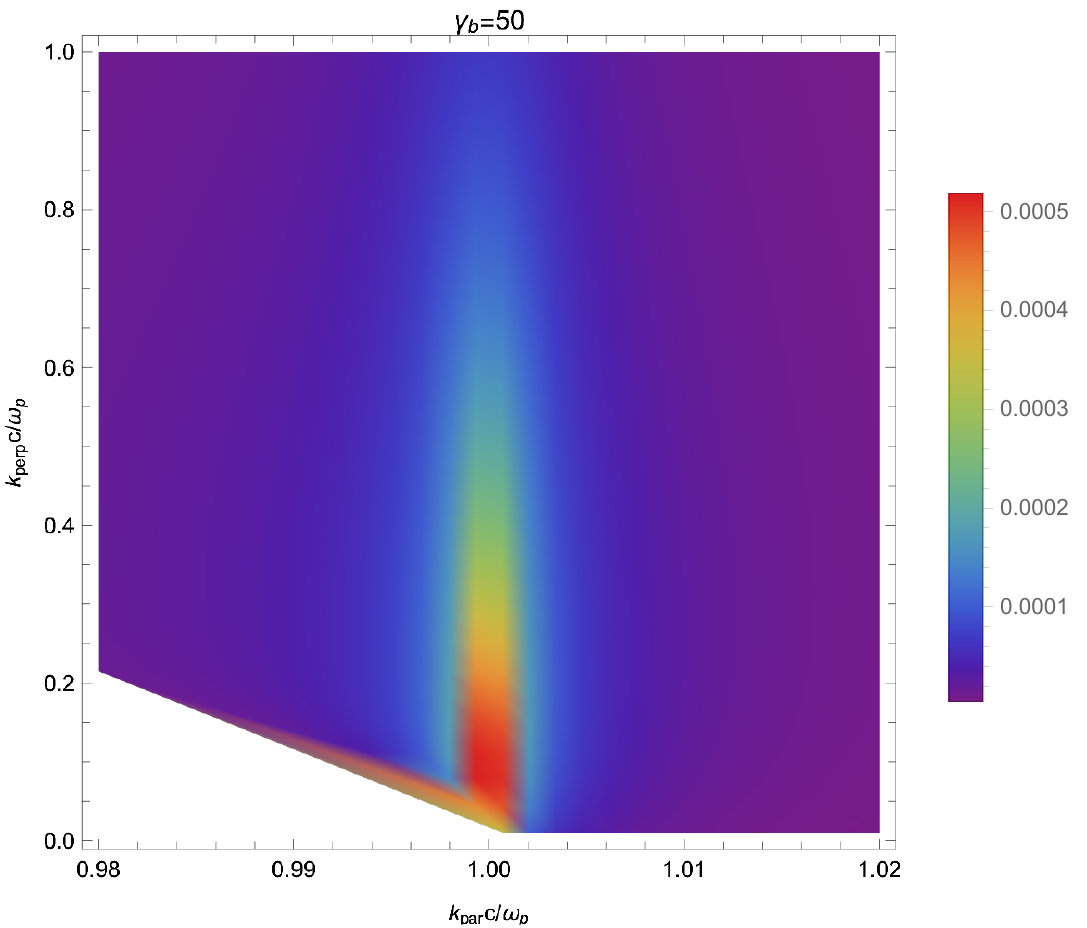}
\includegraphics[width=.3\linewidth]{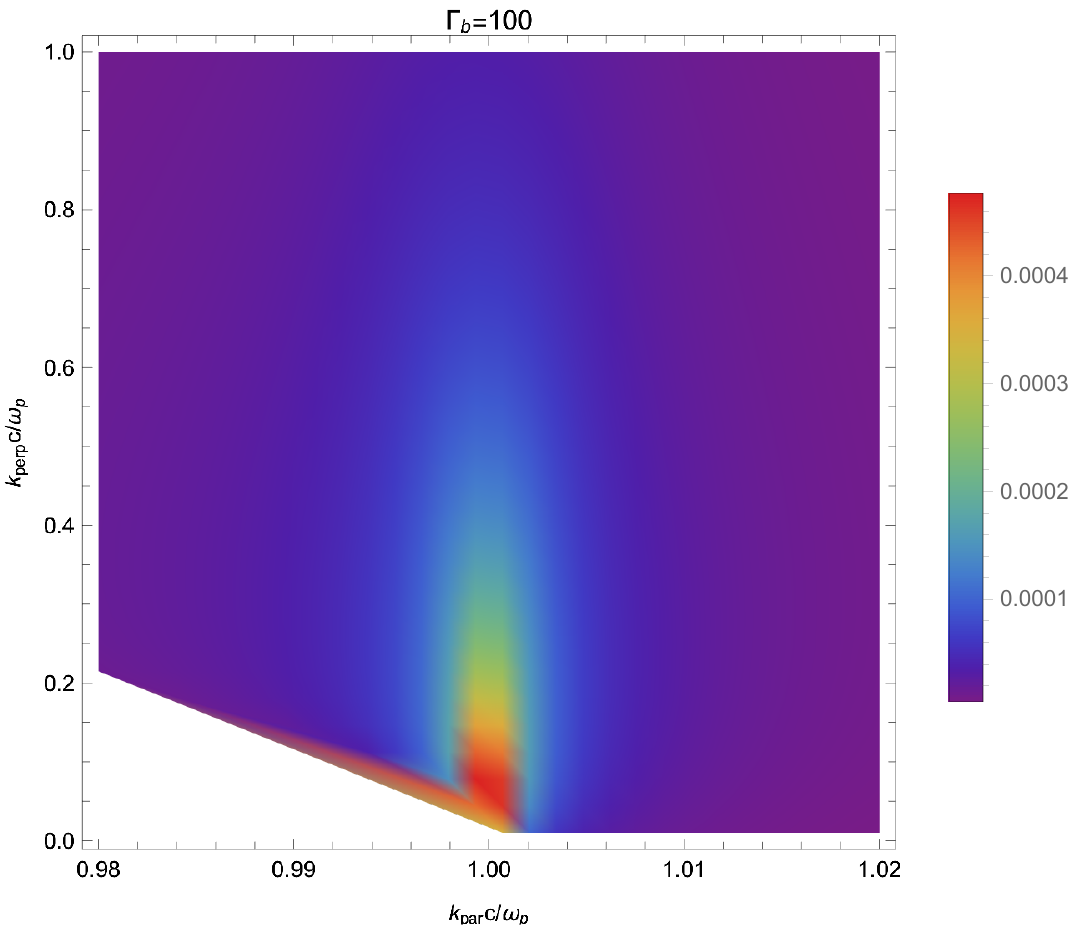}

\caption{Kinetic growth rate $\delta_{k}$ from Eq.~\ref{eq:grkinetic} for oblique modes of instability for a density contrast of $\alpha \sim 10^{-4}$ and beam Lorentz factors of $\gamma =$ 25, 50 and 100. The beam momentum distribution follows Eq.~\ref{eq:2Dnormal}. In the white areas the modes are damped rather than growing. }
\label{fig:grkinetic}
\end{figure}

As the beam undergoes diffusion in momentum space, the momentum spread of the beam increases and the growth of instabilities moves from reactive to kinetic regime. We explore the instability growth for a case without correlations and the corresponding growth rates for a beam-plasma system with a density contrast of $\alpha \simeq 10^{-4}$ and mean momentum characterized by beam Lorentz factors, $\gamma$, of 25, 50 and 100 are shown in Fig. \ref{fig:grkinetic}. The spread of the beam in the forward direction is correspondingly set by $(\gamma/10)m_{e}c$ in each case, which implies that the beam is energetically broad and the modes grow in the kinetic regime.

 \section{Quasilinear relaxation} \label{sec:diffusion}

%After the onset of Langmuir oscillations, the growth of the modes owing to instabilities compete with the change in phase velocities characterised by a quasilinear relaxation phase.
Let us, however, remind ourselves that the quasilinear approach has limited applicability and is only valid when the thermal energy in the plasma is larger than that of the waves undergoing oscillation. During this phase, the momentum spread of the injected beam distribution function increases until the beam becomes stable according to Penrose's criterion, i.e., stability is achieved when the perpendicular momentum spread becomes comparable to the beam Lorentz factor \citep{buschauer1977temperature}. Therefore, even when starting out with a nearly monochromatic beam, or a beam with a small energy spread, for which the instability growth rate is reactive and occurs over a much shorter timescale, as $t_\text{inst} \sim 1/(\delta\omega_\text{p})$, the beam relaxation will broaden the energy width of the beam, causing the instability to proceed in the kinetic regime. Nonlinear effects also play a key role in the fate of the pair beam, as the damping competes with the growth until saturation is reached. The corresponding beam evolution is governed by the generalized Boltzmann equation in the collisionless regime, which reduces to a Fokker-Planck equation:

\begin{equation}
\frac{\partial}{\partial t} f(\mathbf{p}, t)= -\frac{\partial}{\partial p_i}\left[V_i(\mathbf{p}, t) f(\mathbf{p}, t)\right]+\frac{\partial}{\partial p_i}\left[D_{ij}(\mathbf{p}, k, t) \frac{\partial}{\partial p_j} f(\mathbf{p}, t)\right],
\label{eq:fp}
\end{equation}
 
where the first (drift) term represents energy loss due to instability as change of momentum $\mathbf{V}(\mathbf{p}, t) = \dot{\mathbf{p}}$ and the second term characterises diffusion through the diffusion coefficient $D_{ij}$.

For a laboratory plasma, the beam Lorentz factor $\gamma$ is much smaller than that of astrophysical beams, thus the growth rate is substantial and the energy loss through instabilities is comparatively large. Ignoring collisional energy loss at the moment, the spectral energy density $W$ grows via the resonant electrostatic unstable modes,

\begin{equation}
\frac{\partial W}{\partial t}=2\operatorname{Im}(\omega) W.
\label{eq:specedens}
\end{equation}

\subsection{Analytic solution for the Fokker-Planck evolution of a 1D Gaussian} \label{subsec:1Danal}
For the ease of computation, we will assume a one-dimensional system in momentum space with drift and diffusion terms that depend exclusively on time. Thus, in one dimension, the Fokker-Planck equation simplifies to:
\begin{equation}
    \frac{\partial f}{\partial t} = - V(t) \frac{\partial f}{\partial p} + D(t) \frac{\partial^2 f}{\partial p^2}, \label{eq:fp-t}
\end{equation}
where we assume for simplicity that the coefficients $V(t)$ and $D(t)$ do not depend on the momentum $p$.
We use the Ansatz of a Gaussian distribution with a time-dependant mean $\mu(t)$ and width $\sigma(t)$ to solve this partial differential equation,
\begin{equation}
    f(p, t) = \frac{1}{\sqrt{2 \pi} \sigma(t)} \exp\left( - \frac{(p - \mu(t))^2}{2 \sigma(t)^2} \right).
    \label{eq:ini_FP}
\end{equation}
Inserting this Ansatz into Eq.~\ref{eq:fp-t} and dividing by $f$ gives us an equation for the combined evolution of $\sigma$ and $\mu$. Reordering the terms yields
\begin{equation} 
  \left[  \frac{p - \mu(t)}{\sigma(t)^2} \right] \left( \frac{\partial \mu}{\partial t} - V(t) \right) = \left[ \frac{\left( p - \mu(t) \right)^2}{\sigma(t)^2}  - 1\right] \left( \frac{1}{\sigma(t)} \frac{\partial \sigma}{\partial t} - D(t) \frac{1}{\sigma(t)^2} \right).
\end{equation}
Since the terms in square brackets depend on momentum, they cannot vanish in general and therefore the terms in round brackets need to vanish for a solution. This is equivalent to decoupling the evolution of the mean from the standard deviation. We get two independent equations for the evolution of $\mu$ and $\sigma$.
\begin{equation}
    \frac{\partial \mu}{\partial t} = V(t) \Rightarrow \mu(t) = \mu(t=0) + \int_0^t \mathrm{d}t^\prime V(t^\prime), \label{eq:mu_integral}
\end{equation}
\begin{equation}
    \sigma \frac{\partial \sigma}{\partial t} = D(t) \Rightarrow \sigma(t) = \sqrt{\sigma(t=0)^2 + 2 \int_0^t \mathrm{d}t^\prime D(t^\prime)}.
    \label{eq:sigma_integral}
\end{equation}
This treatment can be generalized to two or three dimensions.

The most general field energy density we can construct for the early time regime would consist of a constant term associated with initial noise and an instability term growing exponentially with the rate $2\delta \omega_{p}$,
\begin{equation}
    W(t) = W_0 + W_1 \exp\left( 2\delta \omega_\text{p} t\right) \label{eq:general_energy_density},
\end{equation}
where we can construct the respective drift and diffusion terms as proportional to the derivative of the field energy density and proportional to the field energy density as 
\begin{equation}
    V(t) = V_1\exp\left( 2 \delta \omega_\text{p} t\right), 
\end{equation}
\begin{equation}
    D(t) = D_0 + D_1 \exp\left( 2 \delta \omega_\text{p} t\right).
    \label{eq:driftdiff1D}
\end{equation}
Since the drift term depends on the derivative of the energy density there is no contribution from a constant term.\\
Inserting these into Eq.~\ref{eq:mu_integral} and Eq.~\ref{eq:sigma_integral} will give us the full evolution during the linear growth phase until the saturation regime is reached.
\begin{equation}
    \Delta \mu(t)=\mu(t=0)-\mu(t) = - \frac{V_1}{2\delta\omega_\text{p}}\left[\exp\left( 2 \delta \omega_\text{p} t \right)-1\right],\label{eq:mu-gen}
\end{equation}
\begin{equation}
    \sigma(t) = \sqrt{\sigma(t=0)^2 + 2 D_0 t + \frac{D_1}{\delta\omega_\text{p}}\left[\exp\left( 2\delta \omega_\text{p} t\right)-1\right]} \label{eq:sigma-gen}
\end{equation}
For the mean we find that the entire evolution is described by an exponentially growing shift. For the width, the story is a little more complicated. There can be three different phase of its evolution:
\begin{enumerate}
    \item Initially there will always be a phase where the width remains at its initial value.
    \item Depending on the initial energy density of the electromagnetic fields this can be followed by a phase where the width grows with the square root of time.
    \item At late times the instability term will dominate and the width grows exponentially.
\end{enumerate}
For larger $t$ we find that
\begin{equation}
    \Delta \mu(t) \approx - \frac{V_1}{2\delta \omega_p} \exp\left( 2 \delta \omega_p t\right) ,
\end{equation}
\begin{equation}
    \sigma(t) \approx \sqrt{\frac{D_1}{\delta \omega_p}} \exp\left( \delta \omega_p t\right) .
\end{equation}
We find that deep in the linear regime the shift of the mean momentum grows with the same rate as the energy density of the electromagnetic fields ($2\delta \omega_\text{p}$), whereas the standard deviation grows with the same rate as the amplitude of the unstable modes ($\delta \omega_\text{p}$).

We know that the growth of the field due to the instability can not continue indefinitely but stops when the fields assume a constant value, which however is not instantaneous. We can model this saturation by modifying $\delta \omega_\text{p} t$ as  
\begin{equation}
    \delta \omega_\text{p} t \rightarrow - \log_\kappa \left( \kappa^{-\delta \omega_\text{p} t} + \kappa^{- \delta \omega_\text{p} t_\mathrm{final}} \right) = - \frac{\log \left( \kappa^{-\delta \omega_\text{p} t} + \kappa^{- \delta \omega_\text{p} t_\mathrm{final}} \right)}{\log \kappa}.
\end{equation}
Here $t_\mathrm{final}$ is the time of saturation and $\kappa$ describes how fast this saturation proceeds. For $\kappa \rightarrow \infty$ the saturation is instantaneous.
This substitution gives us the evolution of $\mu$ and $\sigma$ into the non-linear regime,
\begin{equation}
    \Delta \mu(t) =  - \frac{V_1}{\delta\omega_\text{p}} \left( \kappa^{-\delta \omega_\text{p} t} + \kappa^{- \delta \omega_\text{p} t_0}\right)^{-\frac{2}{\log \kappa}} ,\label{eq:mu_full}
\end{equation}
\begin{equation}
    \sigma(t) = \sqrt{\sigma(t=0)^2 + 2 D_0 t + \frac{2 D_1}{\delta\omega_\text{p}} \int_0^t \mathrm{d}t^\prime \left( \kappa^{-\delta \omega_\text{p} t^\prime} + \kappa^{- \delta \omega_\text{p} t_0}\right)^{- \frac{2}{\log \kappa}}} .\label{eq:sigma_full}
\end{equation}
For the mean momentum we find that the shift stops in the nonlinear regime when the derivative of the field energy density vanishes. The width of the distribution evolves as $\sqrt{t}$ with a diffusion coefficient determined by the spectral energy density of the electromagnetic field amplitude.

\subsection{Solution to the Fokker-Planck equation for a 2D Gaussian}
\label{subsec:2Dnum}

For an initial phase-averaged ensemble, the momentum-diffusion tensor is determined by the mode-weighted spectral energy density at resonance \citep{brejzman1974powerful},

\begin{equation}
D_{\alpha \beta}(\mathbf{v})=\pi e^{2} \int  d^{3} k\,W(\mathbf{k}, t) \frac{k_{\alpha} k_{\beta}}{k^{2}} \delta(\mathbf{k} \cdot \mathbf{v}-\omega),
\label{eq:diffcoeff}
\end{equation}
where $\delta$ denotes the Dirac delta function here.
In general the diffusion coefficients depend on beam momenta $\mathbf{p}$ through the velocity $\mathbf{v}=\mathbf{p}/\gamma$ and are also sensitive to the initial beam distribution function through the growth rate in the exponent in the expression for the spectral energy density. Since the angle between the wave vector and the group velocity of particles, $\theta_0$ in the beam is small, the $p$-dependence can approximately be ignored. In this treatment of the momentum diffusion in the beam-plasma system, the delta function associated with the resonance condition in Eq.~\ref{eq:diffcoeff} selects the corresponding fastest growing modes contributing to the growth of the spectral energy density (which carries wavenumber ${\bf k}$ as argument), i.e.,

\begin{equation}
    W(\mathbf{k},t)=W_{0}(\mathbf{k})\exp[ 2 \delta_{\mathbf{k}}\omega_\text{p} t], 
\end{equation}

where $\delta_{\mathbf{k}}$ is the growth rate of the plasma instabilities for mode $\mathbf{k}$, normalised in terms of plasma frequency $\omega_\text{p}$ and $W_{0}(\mathbf{k})$ is the initial spectral energy density in the plasma which could be estimated as the thermal energy density of the background electrons divided by the three-dimensional resonance width.
 
The total energy density of the waves is then given by
\begin{equation}
    W (t) = \int d^3k\, W(\mathbf{k})\sim W_{0} \exp[ 2 \delta_\mathrm{max}\omega_\text{p}t],
\end{equation}
where $\delta_\mathrm{max}$ is the maximum growth rate
and $W_{0} \sim n_\text{p} k_\text{B} T_\text{p}$ is the initial total energy density described by the thermal fluctuations, which can be considered as a noise floor.
In the astrophysical scenario, this would be determined by the temperature of the IGM $T_\text{p}\mathcal{O}(\sim 10\,\mathrm{keV})$. In a laboratory set up, the corresponding temperature is of the order of eVs. 

%The growth of the unstable modes eventually reach a saturation owing to nonlinear effects as well as collisions which must be taken into account in understanding the full scope of collective plasma effects. 

\subsubsection{Drift term}

The first term in the RHS of Eq.~\ref{eq:fp} describes a ``drift'', which essentially represents the energy drain from the beam into the background plasma. As the beam loses energy, the background field energy grows. Energy conservation allows then to estimate the drift coefficient $\upsilon$ in the parallel direction as:

\begin{equation}
V_{\mid \mid}(\mathbf{p}, t) = \dot{p_{\mid \mid}} \sim \delta \omega_\text{p} \frac{W(t)}{n_\text{b}}.
\label{eq:driftpar}
\end{equation}

The above directly follows from the growth of electrostatic field energy due to instability with a growth rate of $\delta$ in units of plasma frequency $\omega_\text{p}$ in the forward direction, i.e., direction of beam propagation. In the transverse direction, the beam momenta is relatively small. Since $p_{\perp} \ll p_{\mid \mid}$, drift in transverse direction is negligible for pair beams designed to mimic astrophysical scenarios in the laboratory.

\subsubsection{Diffusion term}

The diffusion term can be broken down into four components, two diagonal and two off-diagonal contributions. The Cherenkov resonance condition contained in the delta function of the integrand acts as a filtering comb that selects only the fastest growing resonant mode. Thus, the diffusion coefficients in the 2D Cartesian system can be approximated as

\begin{equation}
D_{\alpha\beta}\sim \pi e^2 \frac{W_{0}}{\omega_\text{p}} \exp[2 \delta_{\text{max}} \omega_\text{p} t] J_{\alpha\beta},
\end{equation}
where formally
\begin{equation}
J_{\alpha\beta}=\omega_\text{p}\frac{\int  d^{3} k\,W(\mathbf{k}, t) \frac{k_{\alpha} k_{\beta}}{k^{2}} \delta(\mathbf{k} \cdot \mathbf{v}-\omega)}{\int  d^{3} k\,W(\mathbf{k},t)}.
\end{equation}

The integrals over the resonant modes evaluate to numbers $J_{\mid \mid},J_\perp\sim \mathcal{O}(10)$, and the off-diagonal terms vanish as $J_{\mid \mid, \perp} = J_{\perp, \mid \mid} = 0$.

\subsubsection{Comparison between the drift and diffusion term}

In order to understand the evolution of the pair beam distribution function via the Fokker-Planck equation, it is important to understand which term drives the system. For example, if the drift term is much larger than the diffusion term, the system is advective in momentum space, and the primary consequence of the Fokker-Planck evolution would then be energy loss. In the case of a diffusion term dominating, momentum broadening instead of energy loss would be the main outcome of the pair beam evolution. We now want to estimate the relative impact of the two terms.

In order to do this, at first we consider a comparison along the direction of propagation. The diffusion coefficient, expressed in terms of the spectral energy density of the electromagnetic fields $W(t)$, reads:

\begin{equation}
    D_{\mid \mid} = \pi e^2\frac{W(t)}{\omega_\text{p}} J_{\mid \mid}.
\end{equation}

With the drift coefficient from Eq.~(\ref{eq:driftpar}) 
the dimensionless ratio of the drift to the diffusion term can be estimated as
 
\begin{equation}
    \frac{p_{\mid \mid} \dot{p_{\mid \mid}}}{D_{\mid \mid}} = \frac{p_{\mid \mid}}{\pi e^2 W(t)} \frac{\delta \omega_\text{p}^2 W(t)}{n_\text{b}} = \frac{p_{\mid \mid}}{\pi e^2} \frac{\delta \omega_\text{p}^2}{n_\text{b} J_{\mid \mid}}.
\end{equation}

 Using the definition of plasma frequency, $\omega_{p} = (4 \pi n_\text{p} e^2/m_\text{e})^{1/2}$, and $p_{\mid \mid} = \gamma m_\text{e}$, and noting $J_{\mid \mid}\sim \mathcal{O}(10)$ as noted before,
 
 \begin{equation}
    \frac{p_{\mid \mid} \dot{p_{\mid \mid}}}{D_{\mid \mid}} \approx \delta \frac{\gamma}{\alpha}.
\end{equation}

For standard laboratory plasma conditions with $\gamma = 100$, $\alpha = 10^{-3}$ and a kinetic maximum growth rate, $p_{\mid \mid} \dot{p_{\mid \mid}}/D_{\mid \mid} \sim \mathcal{O}(1-10)$. The transverse momentum $p_{\perp}$ is comparatively smaller than the initial momentum spread, and thus diffusion is the dominant process in the direction perpendicular to propagation.

Independent of the analytical treatment for a 1D Gaussian, shown in Section \ref{subsec:1Danal}, the two-dimensional version of the Fokker-Planck equation, Eq.~\ref{eq:fp}, is numerically solved using an explicit upwind finite difference scheme for a density contrast $\alpha = 10^{-3}$, background plasma temperature $T_\text{p} = 500~\mathrm{eV}$,  plasma density $n_\text{p} = 10^{16} \mathrm{cm}^{-3}$, for an initial injection of a 2D Gaussian symmetric momentum distribution without any correlation assumed between the parallel and perpendicular direction as described in Eq. \ref{eq:2Dnormal}. In order to ensure stability, the size of the timesteps is taken as 1/10 of the maximum resolution set by the corresponding Courant condition.
 
\begin{figure}
\centering
\includegraphics[width=\linewidth]{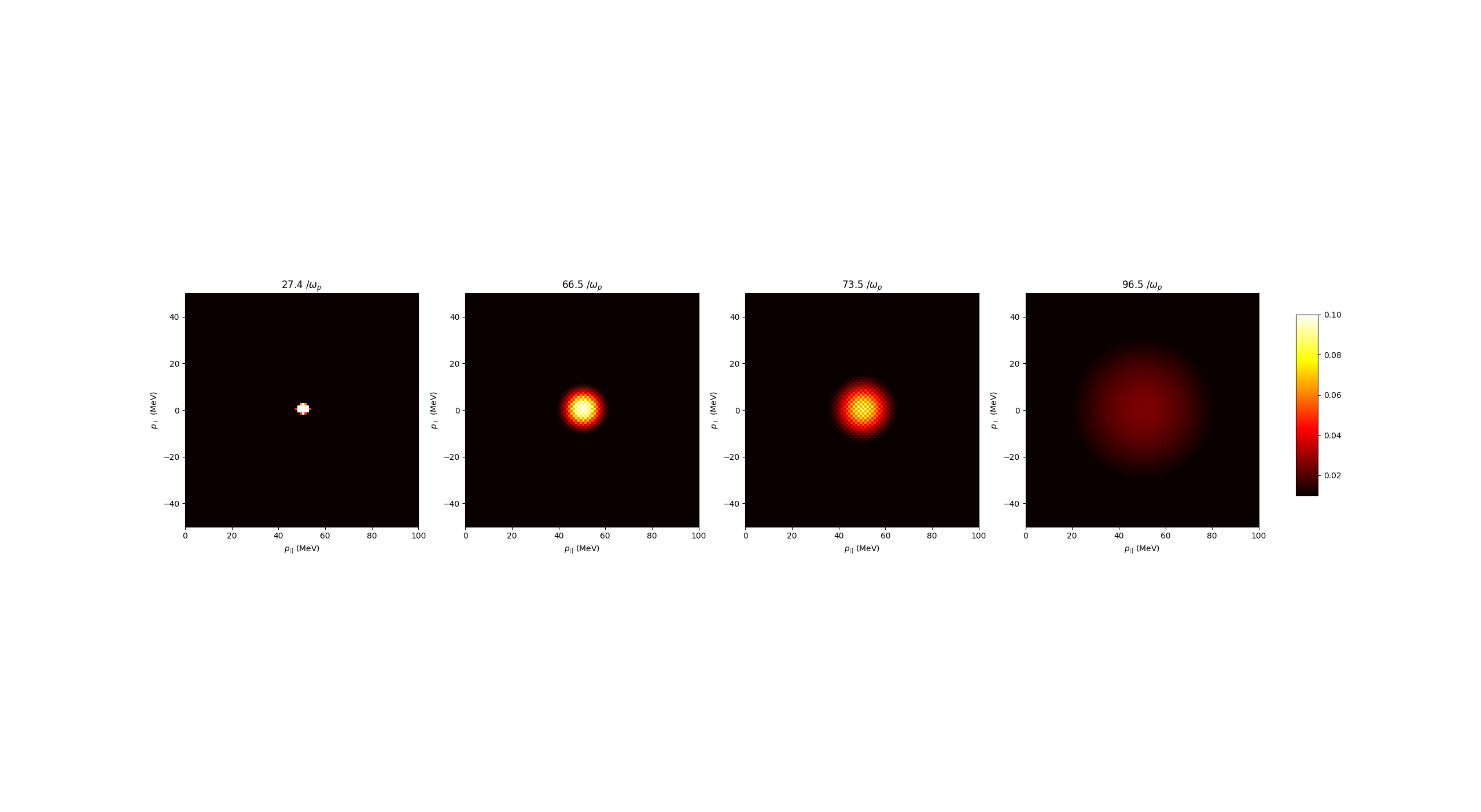}\hfill
\caption{Beam distribution function in $p_{\mid \mid}-p_{\perp}$ plane at various timesteps in the units of inverse plasma frequency. Eq.~\ref{eq:fp} is solved for a plasma with number density $n_{\mathrm{p}}=10^{16}~\mathrm{cm}^{-3}$, density contrast $\alpha \sim 10^{-3}$ and beam Lorentz factors of $\gamma=100$ for initial beam distributions of the form Eq.~\ref{eq:ini_FP}. The background temperature is 500 eV and the color scale has a unit of MeV$^{-2}$. The above panels depict the advective and diffusive Fokker-Planck evolution of the beam through energy loss (negative drift of the mean energy) and a momentum broadening (diffusive spread). The evolution is shown until linear perturbation calculations are valid, i.e., before nonlinear effects become important.
\label{fig:beamevol2D}}
\end{figure}

The input distribution function was a 2D Gaussian with a mean energy of $\gamma = 100$, i.e., 50 MeV at injection. The beam at injection is virtually cold, meaning that the initial width of the beam is very small, which is set by the maximum resolution of the numerical grid. The evolution of the beam distribution function is driven by both the energy loss and diffusive terms in the Fokker-Planck equation, as shown in the numerical solution plotted in Fig. \ref{fig:beamevol2D}. The colourbar shows the probability of finding a beam particle. In Fig. \ref{fig:beamevol2D}, the time evolution of the beam distribution function clearly indicates the shift of the mean beam energy to smaller values representing energy loss, and the broadening of the beam distribution function from an initial narrow Gaussian input whose width is determined by the maximum resolution of the numerical grid, into a wider beam.

\begin{figure}
    \begin{subfigure}[b]{0.5\linewidth}
        \centering
        \includegraphics[width=1\textwidth]{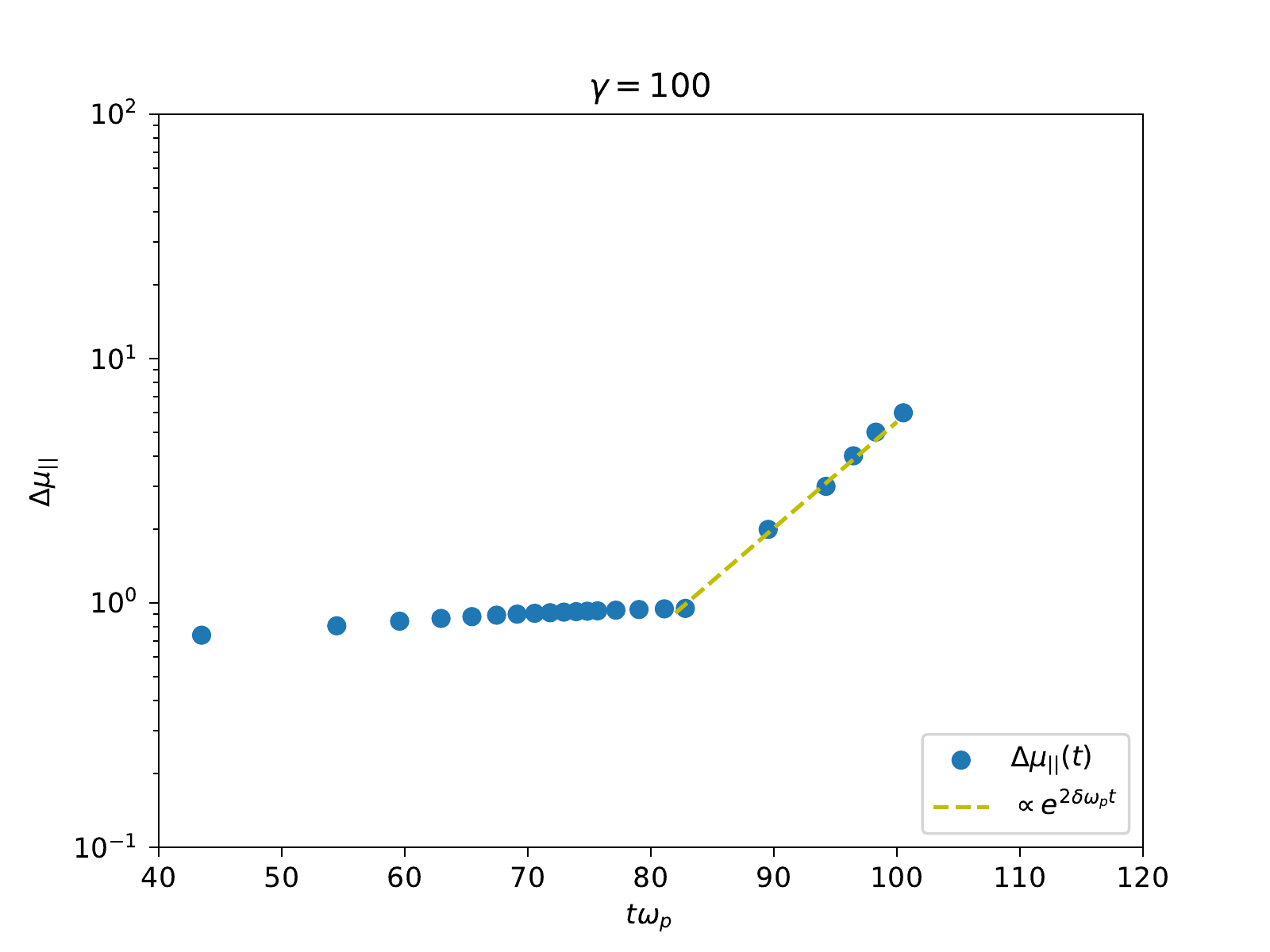}
    \end{subfigure}%
    \begin{subfigure}[b]{0.5\linewidth}
        \centering
        \includegraphics[width=1\textwidth]{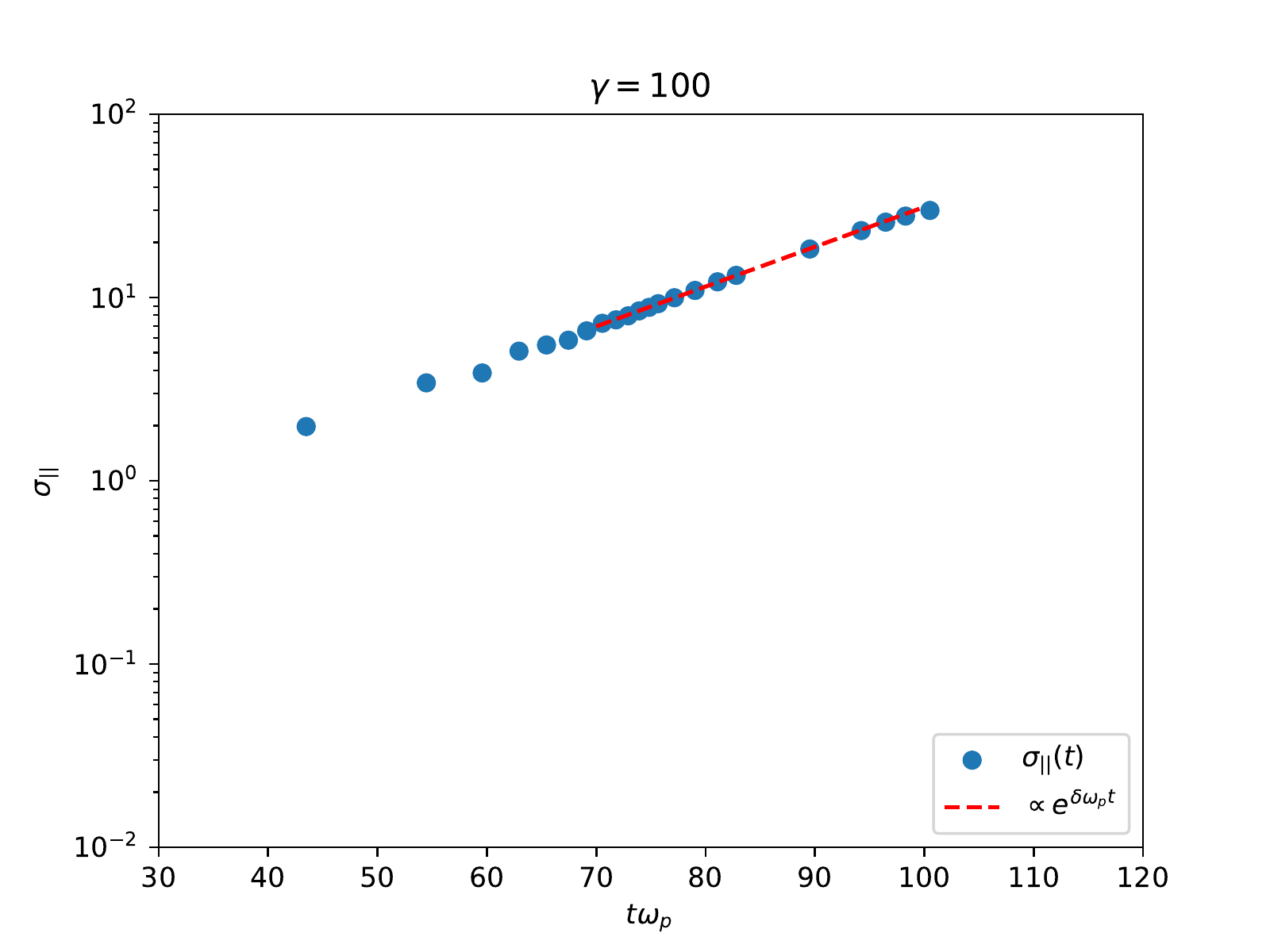}
    \end{subfigure}
    \caption{The left panel shows the time evolution of the drift of the mean parallel momentum of the distribution $\Delta \mu = \left| p_{\|} - \mu_{\|} \right|$. The blue points show the solution of the Fokker-Planck equation, which at late times shows exponential growth as the drift coefficient increases exponentially with time. The green dashed line corresponds to exponential growth at a rate that is twice the growth rate of the field amplitude. The right panel shows the time evolution of the momentum width of the pair beam, in terms of the standard deviation in the parallel direction. The blue points represent the solution of the Fokker-Planck equation. It shows exponential growth at late times as can be expected from the exponentially growing diffusion coefficient. The dashed red line corresponds to exponential growth with the growth rate of the field amplitude.
    \label{fig:mu-sigma-x}}
\end{figure}

The left panel in Fig. \ref{fig:mu-sigma-x} shows the temporal evolution of the absolute value of the shift in the mean momentum in the parallel direction, i.e., $\Delta \mu_{\|} = \left| p_{\|} - \mu_{\|} \right|$. The numerical datapoints are shown in blue, and the dashed green line represents a slope of $\exp(2 \delta \omega_\text{p} t)$. We note that at early times the shift of the mean is not significant, which grows exponentially at later times when the unstable modes have dumped energy from the beam into the background plasma. Similarly, the right panel in Fig. \ref{fig:mu-sigma-x} shows the temporal evolution of the width of the beam in terms of its standard deviation in parallel momentum, $\sigma_{\|}$. The numerical data points are shown in blue, and the red dashed line represents a slope of $\exp(\delta \omega_\text{p} t)$. At early times, the beam width grows as $\sqrt{t}$. Later, at about plasma period of 50 ($t \sim 50/\omega_\text{p}$), we start seeing the exponential growth of the beam width. Thus, the shift of mean energy in the parallel direction  $\Delta \mu_{\|}$ shown in the left panel grows with an exponent twice the instability growth rate, whereas the standard deviation of the beam energy $\sigma_{\|} \sim \sqrt{4D_{\|}t}$ in the parallel direction, grows with an exponent equal to the instability growth rate, corroborating the analytical predictions from Eqs.~\ref{eq:mu-gen} and~\ref{eq:sigma-gen}. This agreement makes a strong case for a unified Fokker-Planck treatment of the evolution of a pair beam. Due to limitations in numerical efficiency, the solution using upwind scheme is restricted to the growth and relaxation phase and does not capture nonlinear effects that lead to instability saturation. In the next section, we delve into PIC simulations of a beam-plasma system in a laboratory setting, in which the onset of nonlinear effects and instability saturation are captured, and compare its findings with the results obtained in this section.

\section{PIC simulations} \label{sec:pic}

%\iffalse
\begin{table}[h]
    %\tiny
    %\small
	\begin{center}
	%\resizebox{.6\textwidth}{!}{
		\begin{tabular}{lr}
		\hline
		\multicolumn{2}{c}{\textbf{Settings of the PIC Simulation}} \\
		\hline
		Number of Dimensions & 2 (x beam direction, y transverse)\\
		\hline
		Boundary conditions in x & periodic \\
		Boundary conditions in y & periodic \\
		Box Length $L_x$ & $500\,c/\omega_\text{p}$ \\
		Box Width $L_y$ & $50\,c/\omega_\text{p}$ \\
		Number of Cells $N_x$ & 4000 \\
		Number of Cells $N_y$ & 400 \\
		Number of particles per cell $N_p$ (per species) & 25 \\
		Timestep $\Delta t$ & 0.95 CFL-Criterion\footnote{The CFL criterion depends on the grid spacing and the field order. For sixth order field interpolation it is given by\\ $\Delta t \leq \frac{120}{149}(\Delta x + \Delta y)\left(\Delta x^2 + \Delta y^2\right)^{-1/2} c^{-1}$} $\approx 0.068 \omega_\text{p}^{-1}$  \\
		Maxwell Solver & Yee \\
		Field Order & 6 \\
		Particle Pusher & Higuera \& Cray \\
		Particle Shape Function & Third Order B-Spline \\
		Current Smoothing & 5-fold (1-2-3-4 steps) \\
		\hline
		Background particles & Electrons \& immobile Protons \\
		Initial Background temperature $T_\text{p}$ & $\left(\gamma - 1\right)\alpha \cdot$ 200\,eV \\
		\hline
		Beam particles & Electrons \& Positrons \\
		Distribution Function & $f(\mathbf{p}; \mu, \sigma_p, \sigma_t) \propto \exp \left( - \frac{\left( p_x - \mu \right)^2}{2 \sigma_p^2} + \frac{p_y^2 + p_z^2}{2 \sigma_t^2}\right)$ \\
		$\mu$ [MeV] & 0.511$\sqrt{\gamma^2 - 1}$\\
		$\sigma_p$ [keV] (alias $\Delta p_x$) & 0.5\\
		$\sigma_t$ [keV] (alias $\Delta p_y$) & 0.5\\
		\hline
		$\alpha$ & 0.001 for $\gamma$ scan\\
		$\gamma$ & 5 for $\alpha$ scan\\
		Total time $T$ & 5000$\,\omega_\text{p}^{-1}$\\
		%\color{red}
		\hline
		\end{tabular}
	%}
	\end{center}
	\caption{Overview of the default PIC Simulation settings for the simulation runs. 
	\label{tab:pic_overview}}
\end{table}
%\fi

Using the public Particle-In-Cell Code EPOCH \citep{Arber:2015hc}, we can validate the predictions of the linear theory and derive further insight on the transfer of energy, the change of the momentum distribution and the saturation of the unstable modes. In the simulations, two Cartesian spatial dimensions, directions parallel (denoted by $x$ below) and perpendicular (denoted by $y$ below) to that of the beam, and three Cartesian dimensions of momentum are resolved in the 2D realization of EPOCH. The boundary conditions of the simulation box are periodic and the size of the simulation grid as well as the number of points are chosen to resolve the instability as derived from the linear calculations. In this case, the dimensions are $L_x = 500 c/\omega_\text{p}$, $L_y = 50 c/\omega_\text{p}$ and the number of grid points is $n_x = 4000$, $n_y = 400$. This resolves each plasma skindepth ($c/\omega_\text{p}$) with 8 cells in each direction leading to a maximum wavenumber of $k_\mathrm{max} = \pi n/L = 8\pi\,\omega_\text{p}/c$ in each direction. The resolution of modes is $\Delta k_x = \pi L_x^{-1} =c\pi/(500\omega_\text{p})$ and $\Delta k_y = \pi L_y^{-1} = c\pi/(50\omega_\text{p})$ respectively. 

The simulation contains four kinds of particle species, beam electrons and beam positrons with a momentum distribution that approximates a cold beam with a Gaussian distribution with a standard deviation of $\sigma = 0.5\,\mathrm{keV}$ and a mean value that is equivalent to the momentum of a particle with Lorentz boost $\gamma$, $\mu_x = (\gamma^2 - 1)^{1/2} m_\text{e} c$, and background electrons and protons with a thermal distribution of $T_\text{p} = 200\,\mathrm{eV} \cdot \left(\gamma - 1\right) \alpha$ at the start of the simulation\footnote{This ensures that the total energy of the background plasma is always the same fraction of the total energy in the simulation. The ratio of initial beam energy to thermal plasma energy is $\epsilon = (n_\text{b}\left(\gamma - 1\right) m_\text{e})/(\frac{3}{2}n_\text{p}k_\text{B} T_\text{p})\approx 1700$.}. The protons have a realistic mass ratio of $m_\text{p} = 1836\,m_\text{e}$ and are assumed to be immobile, i.e., they are not pushed by the particle pusher during the simulation. Each species is resolved by 25 particles per simulation cell, amounting to 40 million total particles per species. The initial momentum distribution is loaded by a sampling-and-rejection method. The density of the background plasma is nominally\footnote{Since all relevant quantities are expressed with respect to the plasma frequency, the absolute value does not influence the simulation result.} set to $n_\mathrm{p} = 10^{16}\,\mathrm{cm^{-3}}$ and the density of the beam particles is respectively determined by the parameter $\alpha = n_\mathrm{b}/n_\mathrm{p}$. Benchmark values of $\gamma = 5 $ and $\alpha = 10^{-3}$ are chosen. 

The simulations use a Yee-Maxwell solver with sixth-order field interpolation and a third-order b-spline shape to deposit the particles onto the grid and a particle pusher as described by \cite{hc_push}. The timestep of the simulations is set to 0.95 of the CLF criterion equalling $0.068\,\omega_\text{p}^{-1}$. In addition, strided current filtering is applied to reduce the influence of numerical instabilities by smoothing the current. Smoothing steps with strides of 1, 2, 3, and 4 cells are applied five times before each particle pushing phase. The filtering follows the scheme described in \cite{vay2014pic}. An overview of the simulation parameters can be found in table \ref{tab:pic_overview}. 

\begin{figure}
    \begin{subfigure}[b]{0.5\linewidth}
        \centering
        \includegraphics[width=1\textwidth]{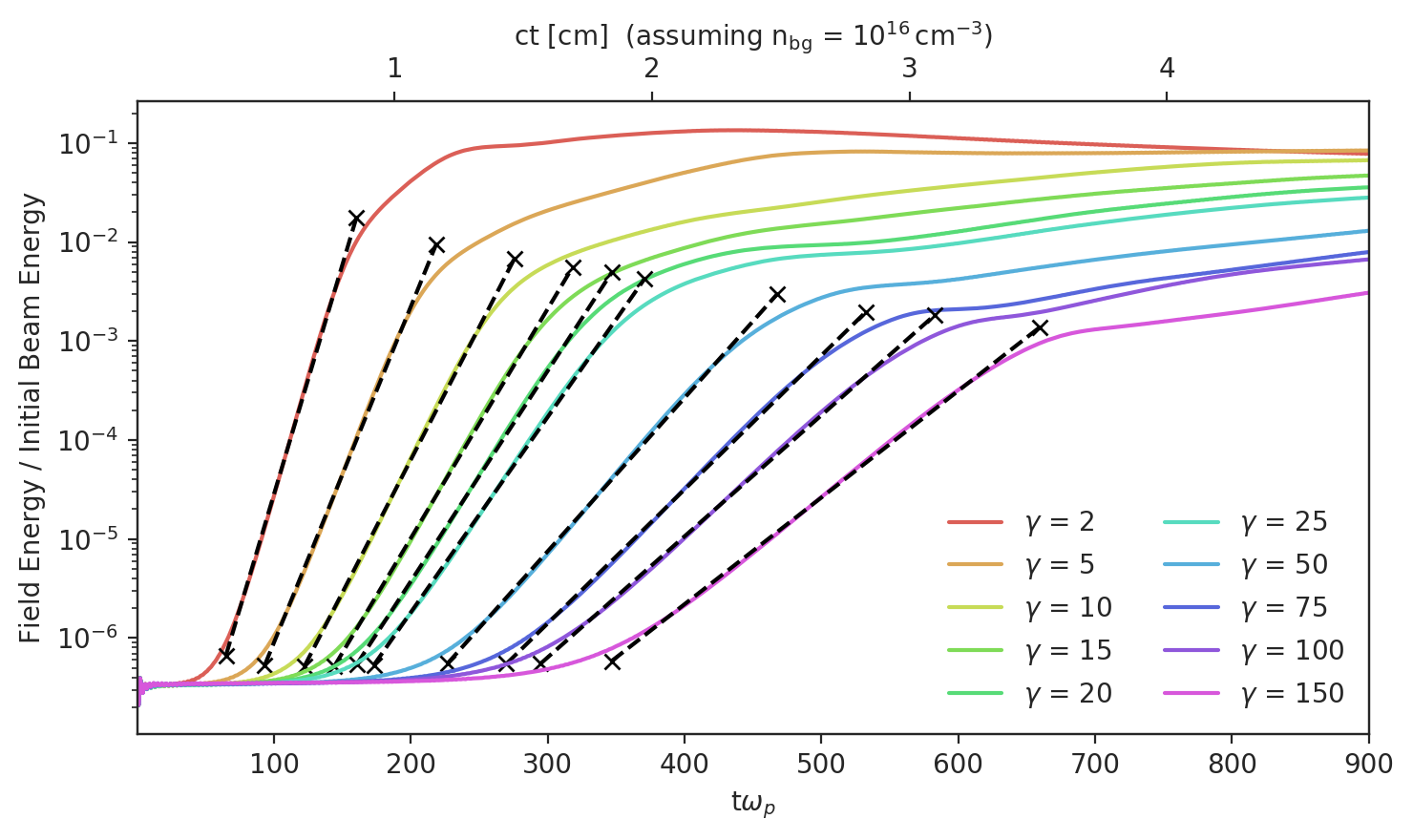}
    \end{subfigure}%
    \begin{subfigure}[b]{0.5\linewidth}
        \centering
        \includegraphics[width=1\textwidth]{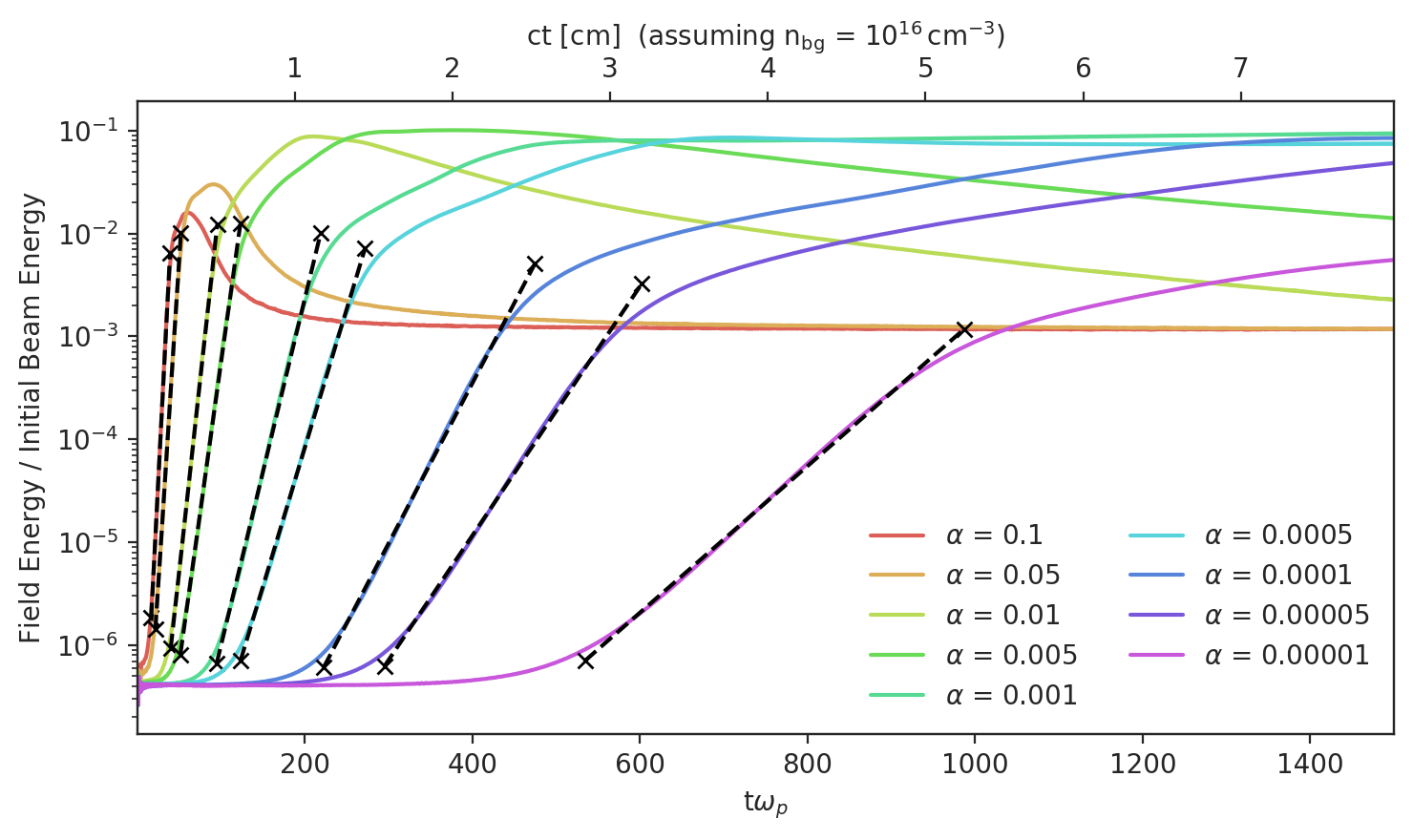}
    \end{subfigure}
    \caption{The energy densities of the electric and magnetic fields normalized to the initial energy density of the neutral pair beam as a function of simulation time for runs with varying $\gamma$ and $\alpha = 10^{-3}$ (left) and varying $\alpha$ and $\gamma = 5$ (right). The simulation runs start with a field energy at a stable noise level. A period of exponential growth, dubbed as the linear growth phase, follows until a saturation regime, where the energy density remains almost constant is reached. The start and end of the linear growth phase is indicated by a cross and the linear growth rate is indicated by a dashed line. The linear growth phase is determined by a fit of the logarithm of the field energy density to equation \ref{eq:ramp_fit}. The $x$-axis is shown in terms of two scalings, the time of simulation runs is shown in units of the inverse plasma frequency as the bottom axis, and the distance travelled by light in real space during this time assuming a background density of $n_\text{p} = 10^{16}\,$cm$^{-3}$ is shown as the top axis. The range of the time axis is chosen such that approximate saturation is reached for the cases shown.}
    \label{fig:energy_overview}
\end{figure}

\begin{figure}[ht] 
  \begin{subfigure}[b]{0.5\linewidth}
    \centering
    \includegraphics[page=2, width=\textwidth]{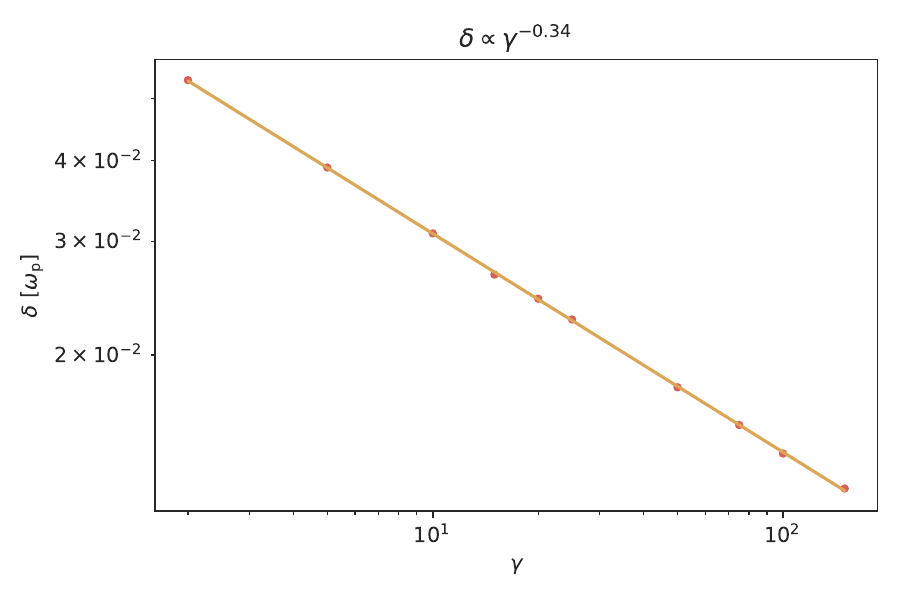}
    %\includegraphics[width=1\textwidth]{gamma_growthrate.png}
    %\caption{For varying $\gamma$ and $\alpha = 0.001$ the power-law was fitted to the data points with a slope of $\delta \propto \gamma^{-0.34}$.} 
  \end{subfigure}%%
  \begin{subfigure}[b]{0.5\linewidth}
    \centering
    \includegraphics[page=2, width=\textwidth]{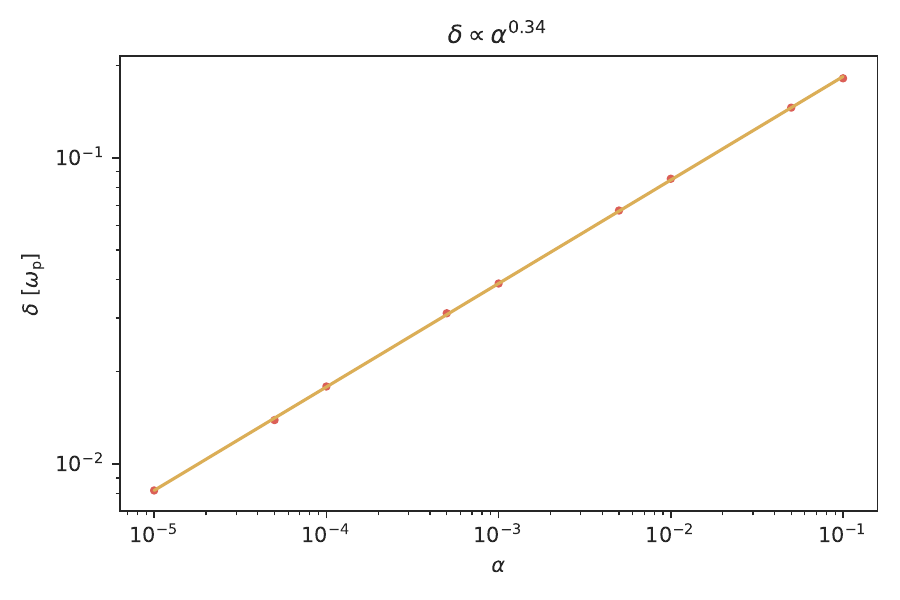}
    %\includegraphics[width=1\textwidth]{alpha_growthrate.png} 
    %\caption{For varying $\alpha$ and $\gamma = 5$ the power-law was fitted to the data points with a slope of $\delta \propto \alpha^{0.35}$.} 
  \end{subfigure} 
  %\caption{The growthrate extracted from different simulation runs. The slopes were extracted from figures \ref{fig:gamma_energy_overview} and \ref{fig:alpha_energy_overview}. }
  \caption{For each simulation run shown in figure \ref{fig:energy_overview} the linear growth rate is extracted by a fit using equation \ref{eq:ramp_fit}. The growth rate is plotted against the parameter that was varied and can be shown to follow a power-law. The left plot shows the growth rate for runs with varying $\gamma$ and $\alpha = 10^{-3}$ where the growth rate is shown to follow a power-law with a spectral index fitted to -0.34. The right plots shows the growth rate for runs with varying $\alpha$ and $\gamma = 5$ and the growth rate follows a power-law with a spectral index fitted to 0.34.
  \label{fig:growthrate}}
\end{figure}

\begin{SCfigure}%[ht]
    \centering
    \includegraphics[page=3, width=0.6\textwidth]{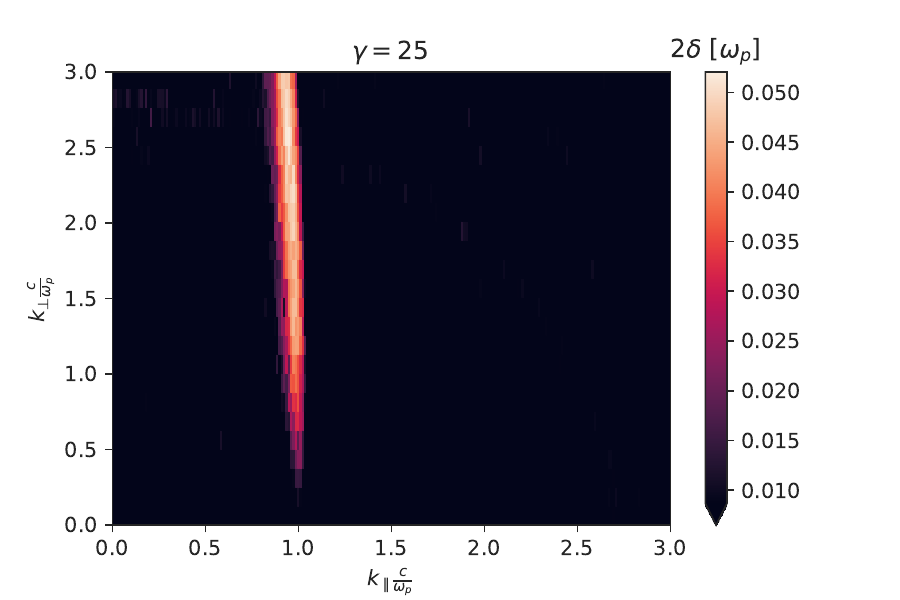}
    \caption{The growth rate as a function of mode $(k_\parallel, k_\bot)$ for a simulation run with $\alpha = 10^{-3}$ and $\gamma = 25$. The growth rate is extracted by Fourier transforming the $E_x$ and $E_y$ fields and calculating the Fourier amplitude parallel to each mode. Then for each mode the growth rate is calculated with a log-linear fit. For calculating these growth rate maps the runs are only considered up to the middle of the (fitted) linear growth phase. The modes with the highest growth rate are found at $k_\parallel \approx \omega_\text{p}/c$. The fastest growing mode is found at an oblique angle. For large $k_\bot$ we can observe that the modes with the largest growth rate shift to slightly smaller $k_\parallel$ values. This effect is closely related to the number of cells used to resolve $\lambda_p$ in transverse direction. In the limit of infinite resolution this feature will vanish.\label{fig:gamma_map}}

\end{SCfigure}

\begin{figure}[ht] 
  \begin{subfigure}[b]{0.5\linewidth}
    \centering
    \includegraphics[page=5, width=\textwidth]{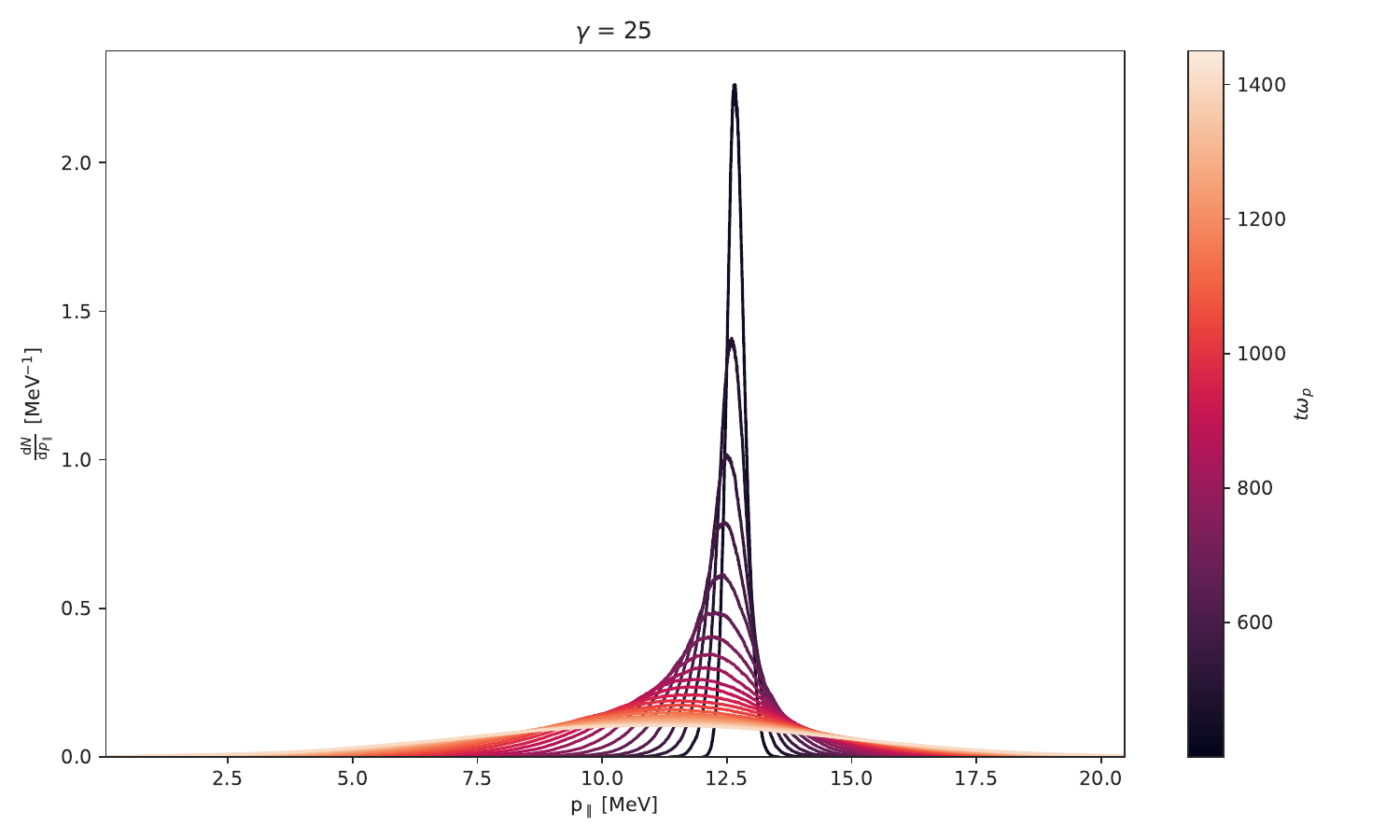}
  \end{subfigure}%%
  \begin{subfigure}[b]{0.5\linewidth}
    \centering
    \includegraphics[page=6, width=\textwidth]{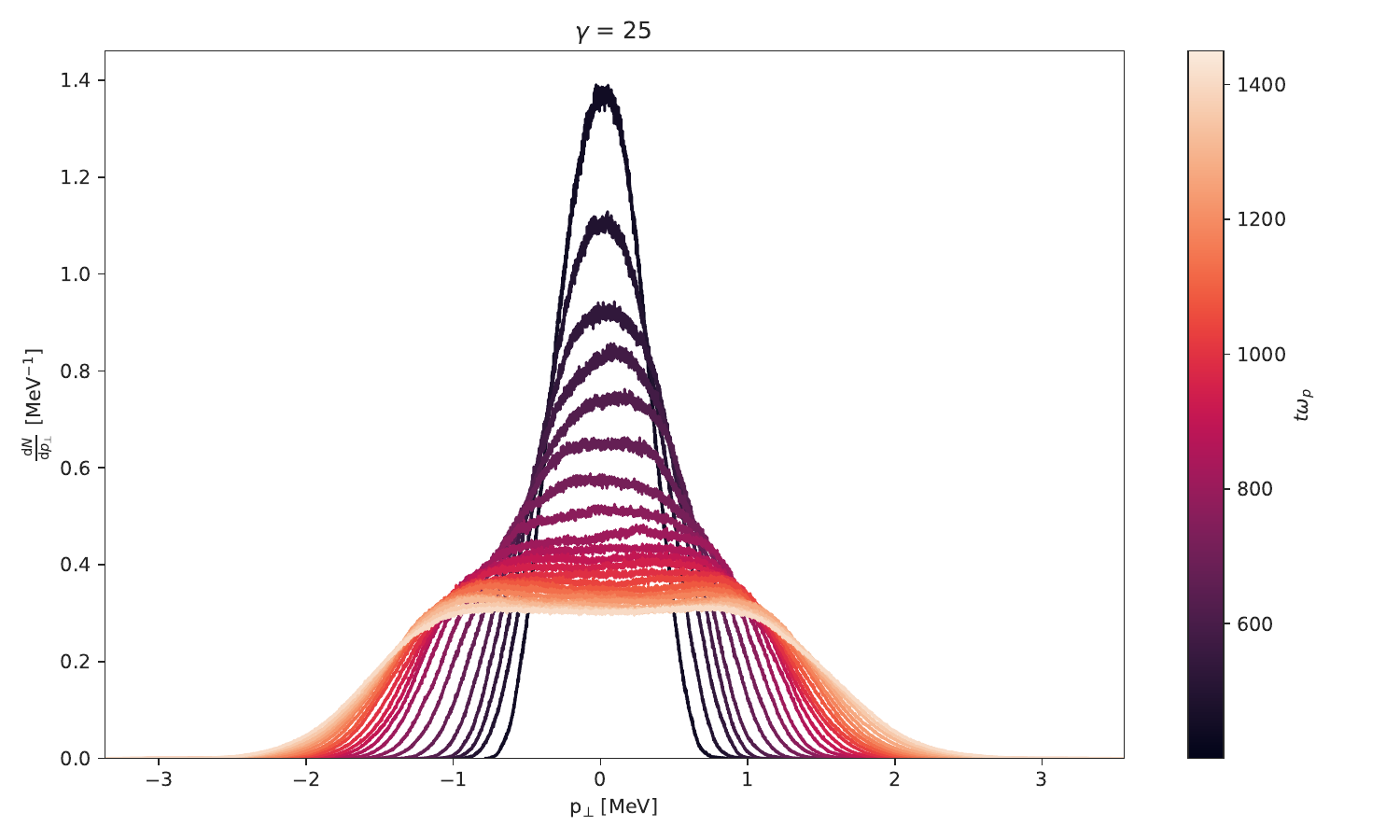}
  \end{subfigure} 
    \caption{Spectrum of $p_\parallel$ (left) and $p_\bot$ (right) taken at regular intervals after linear growth phase starting from a narrow Gaussian initial input for a simulation run with $\gamma = 25$ and $\alpha = 10^{-3}$. The initially cold momentum distribution widens in both directions ($p_\parallel$ and $p_\bot$) during linear growth phase but continues to widen afterwards until a steady state is achieved at a later time. In addition, the mean of the $p_\parallel$ distribution (left) shifts to lower momenta during the simulation run. In comparison with runs with a lower $\gamma$ the final state is wider in both directions but is reached at a later time during the simulation. In comparison with runs with a lower $\alpha$, the final state is wider and is reached faster.
    \label{fig:gamma_p_25}}
\end{figure}

In order to determine the growth rate scaling with the parameters $\gamma$ and $\alpha$, we run multiple simulations where we keep all parameters fixed except one. Figure \ref{fig:energy_overview} shows the total field energies of multiple runs varying $\gamma$ and $\alpha$ respectively. Each simulation run starts with an approximately constant level of noise. The noise level depends on the temperature of the background plasma and the number of macroparticles used to represent it. Once the energy density of the unstable modes exceeds the energy density from the initial shot noise an exponential rise in the total field energy becomes visible. This continues over several $e$-folds of energy growth until eventually a saturation level is reached. It turns out that in the saturation regime the total energy lost by the beam to the electromagnetic fields and the background plasma becomes essentially constant relative to the initial beam energy, as shown in one particular case in Figure \ref{fig:gamma_budget} below.\\
The growth rate in the simulations is defined as the slope of the logarithm of the field amplitude during the exponential growth phase\footnote{The slope of the logarithm of the energy density is twice the slope of the logarithm of the field amplitude, as the energy density is proportional to the square of the field amplitude}. It is extracted from each simulation run by fitting a ramp function (see Eq.~\ref{eq:ramp_fit}, where $2\delta = (c_2 - c_0)/(c_3 - c_1)$ is the growth rate) to the logarithm of the energy density. Aside from the growth rate, this fit also defines the initial noise level ($c_0$), the saturation level ($c_2$), the time of emergence of the linear growth phase ($c_1$), as seen by the energy density of the instability starts to exceed the noise level, and the end of the linear growth phase ($c_3$). In figure \ref{fig:energy_overview}, the beginning and end of the linear growth phase are indicated with a cross for each run and the fitted slope is indicated by a dashed line.
\begin{equation}
    u(t; c) = 
    \begin{dcases}
   c_0        & \text{if } t < c_1 \\
   c_2        & \text{if } t > c_3 \\
   \frac{c_2 - c_0}{c_3 - c_1}\left( t - c_1 \right) + c_0      & \text{otherwise}
  \end{dcases}
.
  \label{eq:ramp_fit}
\end{equation}
From varying the parameters $\alpha$ and $\gamma$ separately, we can roughly establish that the scaling with these parameters (shown in figure  \ref{fig:growthrate}) follows the fastest growing mode in the cold regime described in Eq.~\ref{eq:grreac}, 
\begin{equation}
    \delta \propto \frac{\alpha^{0.34}}{\gamma^{0.34}}.
\end{equation}
Using the information when the linear growth phase begins and ends, we can further investigate the growth rate as a function of the spectral mode $(k_\parallel, k_\bot)$. To this end, we perform a Fourier transform of the electric field components $E_{\|}$ and $E_{\perp}$, the two dimensions resolved by the simulation, and calculate the logarithm of the Fourier amplitude parallel to the wave vector $\mathbf{k}$ for each timestep. The field components are saved in intervals of $5\,\omega_p^{-1}$ for $t < 500\,\omega_p^{-1}$ and in intervals of $50\,\omega_p^{-1}$ otherwise. With a simple linear regression the slope is calculated for each mode. Only timesteps until the middle of the linear growth phase, $t < (c_1 + c_3)/2$, are considered. Figure \ref{fig:gamma_map} shows the growth rate maps for a simulation. Evidently, the dominant modes are situated at $k_\parallel \sim \omega_\text{p}/c$ and a wide spectrum in $k_\bot$. Higher $\gamma$ and lower $\alpha$ parameters lead to a narrower spectrum in $k_\bot$ and thus explain the lower saturation level in these cases.\\
Furthermore, we can observe that in figure \ref{fig:energy_overview} (left) the saturation level of a simulation run decreases with increasing $\gamma$, thus indicating a reduced transfer of energy. The same can be seen in figure \ref{fig:energy_overview} (right) for decreasing $\alpha$ with the caveat that for very large $\alpha$, where $\alpha\gamma$ becomes close to 1, the final field energy density can decrease once again.  For a dilute pair beam the energy lost is divided in equal parts among building up the electric fields and heating the background plasma. However, as shown in figure \ref{fig:gamma_budget} the heating of the background plasma in the non-linear regime can be orders of magnitude larger than the field energy if the beam is only marginally dilute, which could be the case in a laboratory experiment, where the diluteness criterion is $\alpha\gamma \ll 1$. \\
The energy loss is not the same for each beam particle, rather the momentum spectrum is drastically changed from the initially cold distribution. At late times, the distribution is significantly wider and this widening effect continues to alter the momentum distribution long after the end of the linear growth phase. Figures \ref{fig:gamma_p_25} show widening in $p_\parallel$ and $p_\bot$ after the end of the linear growth phase for a simulation run. The widening of the momentum spectrum in $p_\parallel$ also means that some particles even gain energy due to the diffusive relaxation driven by the instability. Moreover, the gain in transverse momentum $p_\bot$ leads to a spatially broader beam in the transverse direction. At a later time, the momentum distribution reaches a steady state of finite width. The final width increases with a higher Lorentz boost $\gamma$ and decreases with a higher density ratio $\alpha$. \\

\begin{SCfigure}
    \centering
    \includegraphics[page=4, width=0.6\textwidth]{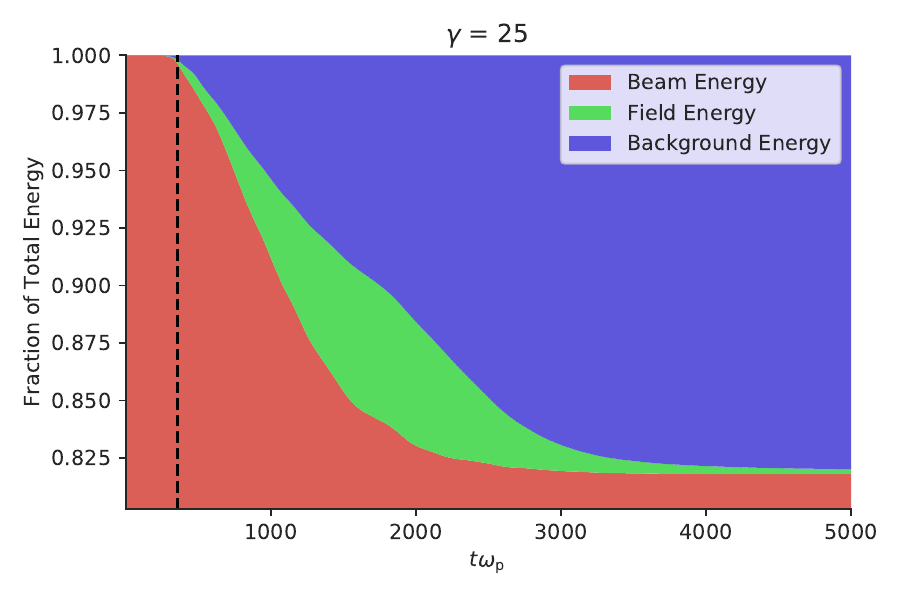}
    \caption{The fraction of the total energy distributed on the kinetic energy of the beam particles, the energy of electromagnetic fields and the kinetic energy of the background particles is shown as it evolves with time for a simulation run with $\alpha = 10^{-3}$ and $\gamma = 25$. The dashed line indicates the end of the (fitted) linear growth phase. Initially, nearly the entire energy of the system is contained as the kinetic energy of the beam particles. Energy is transferred to the fields and the background plasma during and after the linear growth phase. The energy gain of the background plasma can exceed the energy gain of the field by orders of magnitude if the beam is not dilute enough. In the limit $\alpha\gamma \rightarrow 0$ the field energy and background plasma kinetic energy reach equipartition. We note that in an astrophysical environment the beam would be dilute enough that at all times almost the entire energy of the system is concentrated in the background plasma.
    \label{fig:gamma_budget}}
\end{SCfigure}

\begin{figure}
    \begin{subfigure}[b]{0.5\linewidth}
        \centering
        \includegraphics[page=7, width=\textwidth]{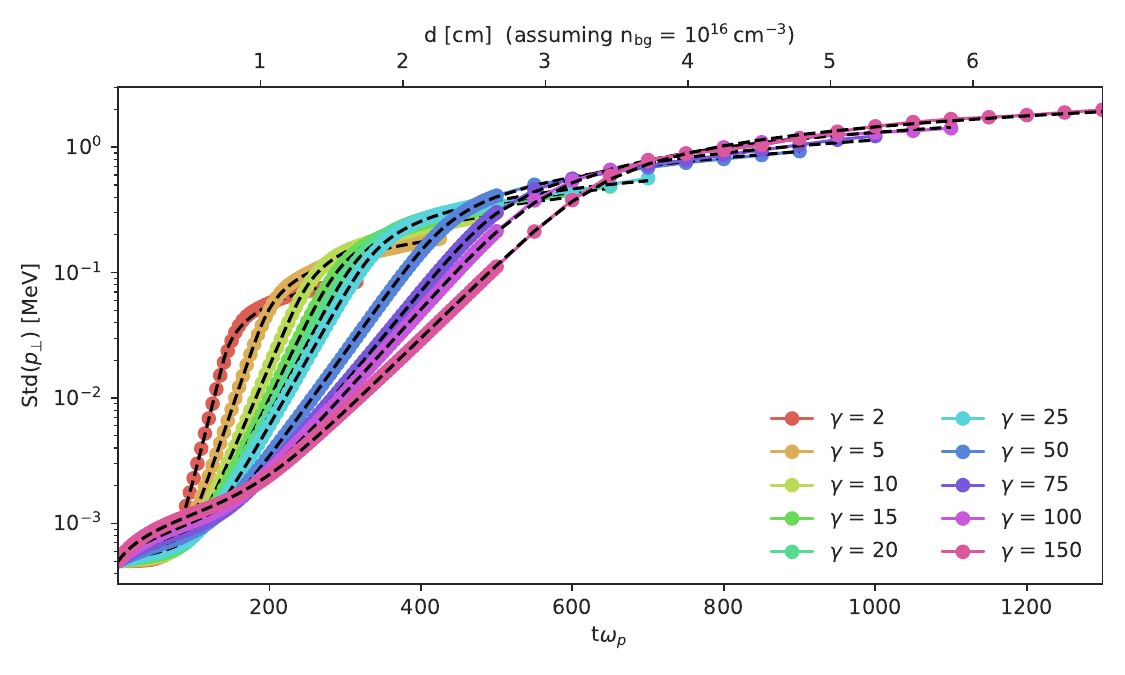}
    \end{subfigure}%
    \begin{subfigure}[b]{0.5\linewidth}
        \centering
        \includegraphics[page=3, width=\textwidth]{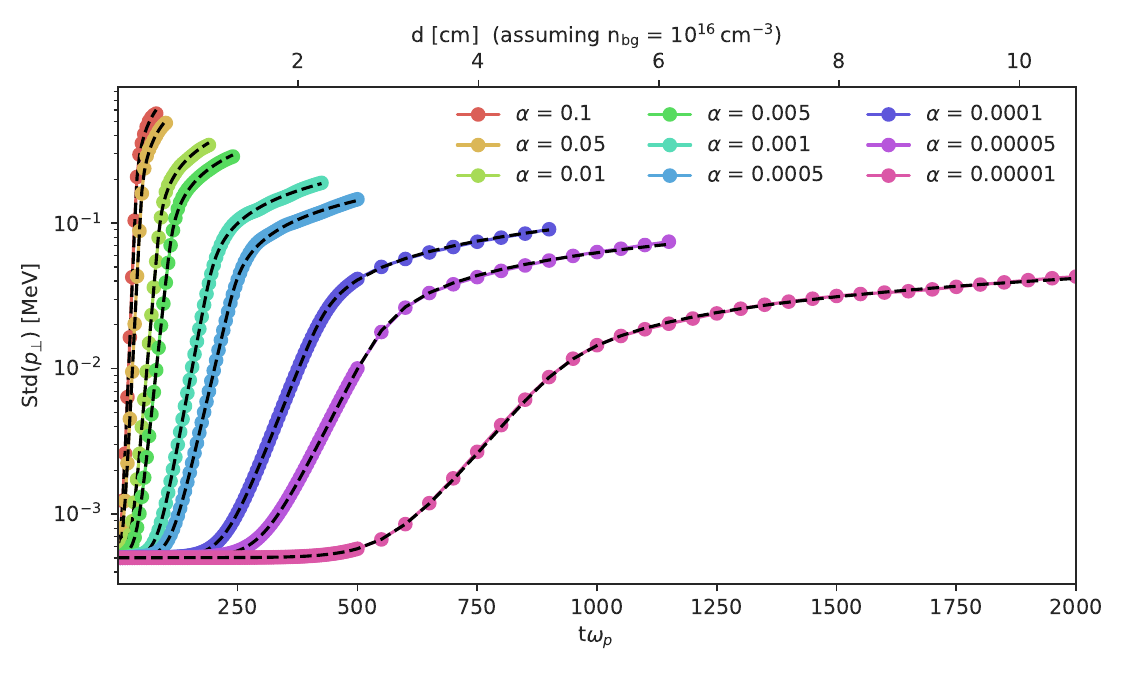}
    \end{subfigure}
    \caption{Evolution of Std($p_\bot$) for simulation runs with varying $\gamma$ and $\alpha = 10^{-3}$ (left) and varying $\alpha$ and $\gamma = 5$ (right). Similar to the the field energy density in Fig. \ref{fig:energy_overview} there is a period of exponential growth followed by a saturation regime. For each run Eq.~\ref{eq:sigma_full} was fitted from the beginning of the simulation until well into the saturation regime. For a high Lorentz boost $\gamma$ at very early times, before the linear growth phase sets in, we can clearly see the $\sqrt{t}$ time scaling associated with the large initial noise. The full list of fitted parameters can be found in table \ref{tab:py_params}.
    \label{fig:evolution_std_py}}
\end{figure}

\begin{figure}
    \begin{subfigure}[b]{0.5\linewidth}
        \centering
        \includegraphics[page=8, width=\textwidth]{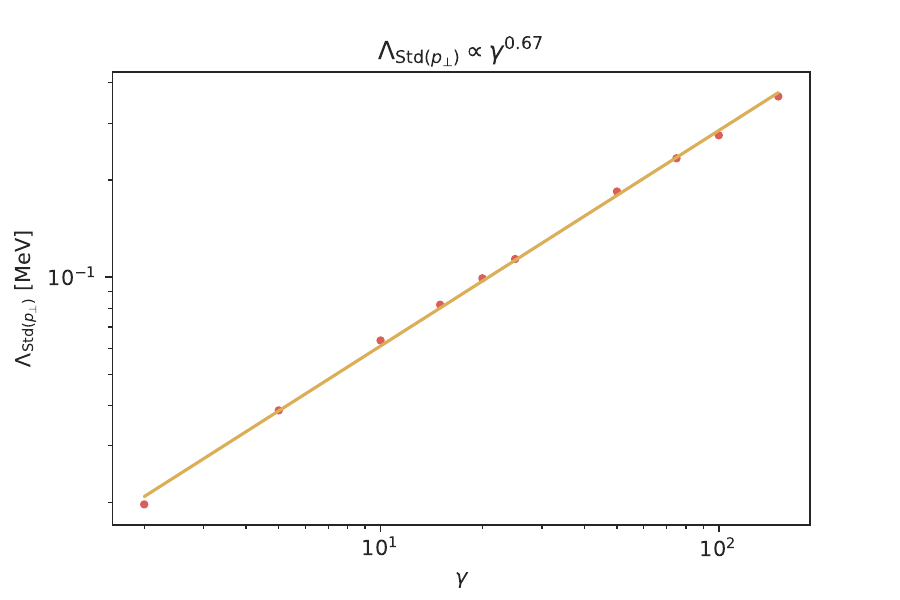}
    \end{subfigure}%
    \begin{subfigure}[b]{0.5\linewidth}
        \centering
        \includegraphics[page=4, width=\textwidth]{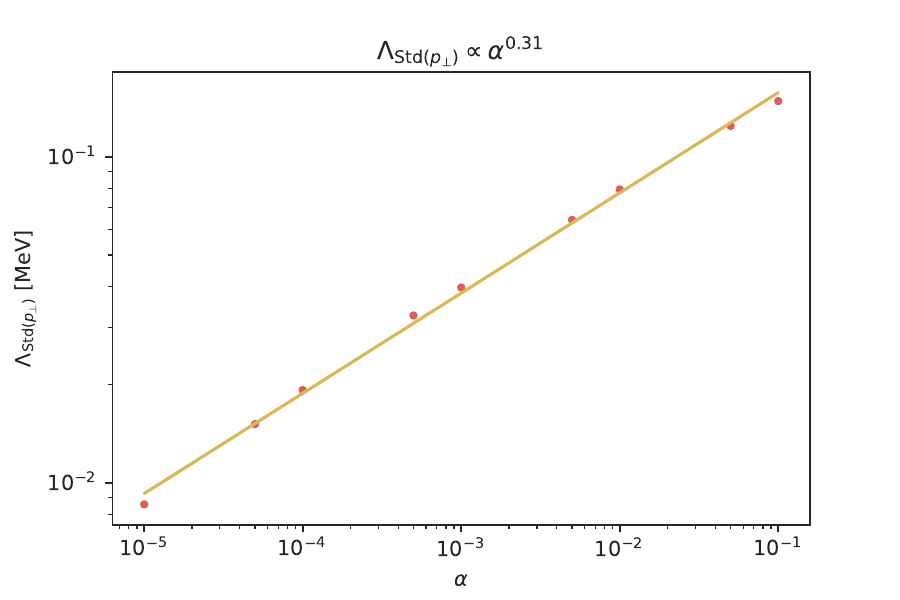}
    \end{subfigure}
    \caption{The transverse momentum spread at the end of the linear growth phase $\Lambda_\mathrm{Std\left( p_\bot \right)} = \sigma_\bot\left( t_\mathrm{final}\right)$ is shown as a function of varying $\gamma$ for $\alpha = 10^{-3}$ (left) and varying $\alpha$ for $\gamma = 5$ (right). The values were obtained by the fit shown in figure \ref{fig:evolution_std_py}. The saturation limit follows a power law for both variables fitted to $\Lambda_{\mathrm{Std}\left( p_\bot \right)} \propto \gamma^{0.66}\alpha^{0.31}$.
    \label{fig:saturation_std_y}}
\end{figure}

In the linear and quasilinear regime, the standard deviations of the momentum distribution in $p_\parallel$ and $p_\bot$ represent the width in the respective directions. Figure \ref{fig:evolution_std_py} shows the evolution of $\mathrm{Std}\left( p_\bot \right)$ for different $\gamma$ and $\alpha$ values. We note that in the PIC simulations the momentum distribution is not necessarily Gaussian at every step in time and thus we denote the standard deviation with Std instead of $\sigma$ which we used to the denote the standard deviation of the Gaussian distribution during the treatment with the Fokker-Planck equation. The same nomenclature is used to distinguish the calculated mean of an arbitrary distribution as Mean instead of $\mu$. The transverse component of the momentum distribution widens exponentially during the linear growth phase of the instability. Before the onset of the linear growth phase, we either find a constant phase or, in case of high $\gamma$, a phase of the width growing as $\sqrt{t}$. The presence of this $\sqrt{t}$ phase depends on the amount of initial noise in the system. Here we find that it is only present for simulation runs with larger $\gamma$, where the temperature of the background plasma is highest, because we kept the parameter $\epsilon$ fixed. By fitting equation \ref{eq:sigma_full} to the standard deviation as function of time we can then extract the fit parameters, such as the rate ($\delta_\mathrm{Std\left(p_\bot\right)}$), the time of saturation $t_\mathrm{final}$, and the saturation level ($\Lambda_\mathrm{Std\left(p_\bot\right)} = \mathrm{Std}\left( p_\bot\right)\left( t_\mathrm{final} \right)$) of the linear instability phase\footnote{Here $\delta_i$ describes the exponential growth rate of the quantity $i$, and $\Lambda_i$ describes the level quantity $i$ reaches at the end of this exponential growth phase.} (see figure \ref{fig:saturation_std_y}). Since at late times nonlinear effects start to become relevant, that are not captured by the analytical solution of the Fokker-Planck equation, we only consider data points up to twice the saturation time denoted in figure \ref{fig:energy_overview}. A full list of all fitted parameters can be found in table \ref{tab:py_params}. \\
Using the scans in $\alpha$ and $\gamma$ we can conclude that both $\delta_\mathrm{Std\left(p_\bot\right)}$ and $\Lambda_\mathrm{Std\left(p_\bot\right)}$ (figure \ref{fig:saturation_std_y}) follow a power-law, where the spectral index of the widening rate is identical to the spectral index of instability amplitude growth rate, as we can expect from the analytical and numerical solutions of Eq.~\ref{eq:fp}, see in particular Eq.~\ref{eq:sigma-gen}, 
\begin{equation}
    \delta_{\mathrm{Std}\left( p_\bot \right)} \propto \alpha^{0.33}\gamma^{-0.33},
\end{equation}
\begin{equation}
    \Lambda_{\mathrm{Std}\left( p_\bot \right)} \propto \alpha^{0.31} \gamma^{0.67}.
\end{equation}

In a similar vein we define the growth rate and saturation level for $\mathrm{Std}\left(p_\parallel\right)$ where $\Delta \mathrm{Mean}\left(p_\parallel\right) = \mathrm{Mean}\left(p_\parallel, t=0\right) - \mathrm{Mean}\left(p_\parallel, t\right)$. We find that, in accordance with our analytical estimate and the numerical solution of the Fokker-Planck equation, Eqs.~\ref{eq:mu_full} and~\ref{eq:sigma_full}, the growth rate of the standard deviation of both $p_\parallel$ and $p_\bot$ grows exponentially at the growth rate of the electric field amplitude, which is half of the growth rate of the electric field energy density. The growth rate of change of the mean value of $p_\parallel$ on the other hand grows exponentially at twice the growth rate of the standard deviations, and thus follows the growth rate of the electric field energy density. All of these processes take place concurrently. The scaling of $\delta$ and $\Lambda$ with $\alpha$ and $\gamma$ is the same for both directions of widening and the shift of mean parallel momentum. Noteworthy is that the scaling of $\Lambda_\mathrm{Std\left(p_\parallel\right)}$ is approximately equal to $\gamma\delta$. It is important to mention that the broadening of the momentum distribution does not stop at the end of the linear growth phase, rather continues with a constant diffusion constant, leading to $\sqrt{t}$ of the standard deviation, for some time. Eventually the energy of the unstable modes is dissipated to other modes and the background plasma and the momentum broadening essentially comes to a halt. For $\gamma > 50$ and $\alpha = 10^{-3}$, the transverse momentum spread will exceed multiple MeVs.

Using the mean of the $p_\parallel$ and the standard deviation of the $p_\bot$ distribution we can also estimate the opening angle $\left< \theta \right>$ of the pair beam at any given time step, 
\begin{equation}
    \left< \theta \right> \approx \frac{\mathrm{Std}\left(p_\bot\right)}{\mathrm{Mean}\left(p_\parallel\right)} \approx \frac{\mathrm{Std}\left(p_\bot\right)}{\gamma m_e}. 
\end{equation}
The opening angle shows the same exponential increase as the field amplitude and the standard deviation of $p_\parallel$ during the linear growth phase. We can see that $\Lambda_{\left<\theta\right>}$ scales a bit stronger with $\gamma$ than $\gamma^{-1}\Lambda_\mathrm{Std\left( p_\bot \right)}$, because the relative loss of mean momentum in beam direction is reduced with higher Lorentz boost. 
\begin{equation}
    \delta_{\left<\theta\right>} \propto \alpha^{0.33}\gamma^{-0.33}, 
\end{equation}
\begin{equation}
    \Lambda_{\left<\theta\right>} \propto \alpha^{0.31}\gamma^{-0.36}. 
\end{equation}
We can give an expression for the opening angle at saturation that can be used to scale it to astrophysical parameters. This does not include additional broadening in the nonlinear regime that in our PIC simulation amounts to an increase by roughly one order of magnitude. 
\begin{equation}
    \Lambda_{\left<\theta\right>} = \frac{1}{4}\,\mathrm{rad} \left( \frac{\alpha}{\gamma}\right)^\frac{1}{3}.\label{eq:theta_sat}
\end{equation}

\begin{table}
    %\centering
    \begin{center}
    \begin{tabular}{c|c||c|c|c|c|c||c}
        $\gamma$ & $\alpha$ & $\delta$ [$\omega_\text{p}$] & $D_0$ [MeV$^2\omega_\text{p}$] & $D_1$ [MeV$^2\omega_\text{p}$] & $t_\mathrm{final}$ [$\omega_\text{p}^{-1}$] & $\kappa$ & $\Lambda_\mathrm{Std\left(p_\bot\right)}$ [MeV] \\
        \hline \hline
2 & 0.001 & $5.55 \cdot 10^{-2}$ & $3.99 \cdot 10^{-11}$ & $4.15 \cdot 10^{-12}$ & 139 & $1.08 \cdot 10^{-24}$ & $1.97 \cdot 10^{-2}$ \\
5 & 0.001 & $4.11 \cdot 10^{-2}$ & $1.06 \cdot 10^{-10}$ & $1.14 \cdot 10^{-11}$ & 190 & $2.26 \cdot 10^{-2}$ & $3.86 \cdot 10^{-2}$ \\
10 & 0.001 & $3.27 \cdot 10^{-2}$ & $2.16 \cdot 10^{-10}$ & $2.16 \cdot 10^{-11}$ & 241 & $2.97 \cdot 10^{-2}$ & $6.36 \cdot 10^{-2}$ \\
15 & 0.001 & $2.85 \cdot 10^{-2}$ & $3.53 \cdot 10^{-10}$ & $3.04 \cdot 10^{-11}$ & 278 & $7.13 \cdot 10^{-2}$ & $8.21 \cdot 10^{-2}$ \\
20 & 0.001 & $2.62 \cdot 10^{-2}$ & $4.74 \cdot 10^{-10}$ & $3.97 \cdot 10^{-11}$ & 302 & $4.20 \cdot 10^{-2}$ & $9.91 \cdot 10^{-2}$ \\
25 & 0.001 & $2.44 \cdot 10^{-2}$ & $6.22 \cdot 10^{-10}$ & $5.14 \cdot 10^{-11}$ & 324 & $4.21 \cdot 10^{-2}$ & 0.114 \\
50 & 0.001 & $1.94 \cdot 10^{-2}$ & $1.35 \cdot 10^{-9}$ & $9.21 \cdot 10^{-11}$ & 413 & $8.56 \cdot 10^{-2}$ & 0.184 \\
75 & 0.001 & $1.69 \cdot 10^{-2}$ & $2.10 \cdot 10^{-9}$ & $1.16 \cdot 10^{-10}$ & 478 & $9.84 \cdot 10^{-2}$ & 0.233 \\
100 & 0.001 & $1.55 \cdot 10^{-2}$ & $2.96 \cdot 10^{-9}$ & $1.68 \cdot 10^{-10}$ & 522 & 0.176 & 0.275 \\
150 & 0.001 & $1.36 \cdot 10^{-2}$ & $4.62 \cdot 10^{-9}$ & $2.33 \cdot 10^{-10}$ & 598 & 0.146 & 0.363 \\
\hline \hline 
5 & 0.1 & 0.182 & $1.13 \cdot 10^{-8}$ & $4.97 \cdot 10^{-9}$ & 37.4 & $1.12 \cdot 10^{41}$ & 0.148 \\
5 & 0.05 & 0.148 & $7.94 \cdot 10^{-9}$ & $1.78 \cdot 10^{-9}$ & 47.4 & $8.90 \cdot 10^{40}$ & 0.124 \\
5 & 0.01 & $8.84 \cdot 10^{-2}$ & $1.77 \cdot 10^{-9}$ & $1.95 \cdot 10^{-10}$ & 84.3 & $1.44 \cdot 10^{4}$ & $7.95 \cdot 10^{-2}$ \\
5 & 0.005 & $7.03 \cdot 10^{-2}$ & $8.51 \cdot 10^{-10}$ & $8.64 \cdot 10^{-11}$ & 107 & 152 & $6.40 \cdot 10^{-2}$ \\
5 & 0.001 & $4.09 \cdot 10^{-2}$ & $1.13 \cdot 10^{-10}$ & $1.30 \cdot 10^{-11}$ & 189 & 86.2 & $3.97 \cdot 10^{-2}$ \\
5 & 0.0005 & $3.26 \cdot 10^{-2}$ & $5.48 \cdot 10^{-11}$ & $6.32 \cdot 10^{-12}$ & 239 & 216 & $3.26 \cdot 10^{-2}$ \\
5 & 0.0001 & $1.90 \cdot 10^{-2}$ & $7.91 \cdot 10^{-12}$ & $1.12 \cdot 10^{-12}$ & 414 & 51.2 & $1.93 \cdot 10^{-2}$ \\
5 & 0.00005 & $1.49 \cdot 10^{-2}$ & $4.34 \cdot 10^{-12}$ & $5.21 \cdot 10^{-13}$ & 532 & 29.7 & $1.51 \cdot 10^{-2}$ \\
5 & 0.00001 & $8.75 \cdot 10^{-3}$ & $7.37 \cdot 10^{-13}$ & $1.14 \cdot 10^{-13}$ & 897 & 25.7 & $8.59 \cdot 10^{-3}$ \\
    \end{tabular}   
    \end{center}
    \caption{The fit parameters ($\delta$, $D_0$, $D_1$, $t_\mathrm{final}$ and $\kappa$) of Eq.~\ref{eq:sigma_full} for Std($p_\bot$) from simulation runs with various values for $\alpha$ and $\gamma$ (all shown in figure \ref{fig:evolution_std_py}). The derived quantity $\Lambda_\mathrm{Std\left(p_\bot\right)}$ is the value of Std($p_\bot$) at time $t_\mathrm{final}$.
    \label{tab:py_params} }
\end{table}

\section{Discussions} \label{sec:discussions}

Beam-plasma systems feature a plethora of microscopic phenomena, and it is only through the lens of collective plasma effects, we can describe the evolution of such systems. In an unmagnetized plasma, a beam propagating through the plasma induces Langmuir oscillations with the resonance condition associated with such electrostatic Cherenkov instability as $\omega - \mathbf{k}\cdot\mathbf{v}=0$. When the initial injection of the beam is monochromatic, the instability is nearly instantaneous (reactive), and its growth rate can be estimated using linear stability analysis hydrodynamically. When the velocity or momentum spread of the beam $\Delta \mathbf{v}$ is large, $\mid \mathbf{k}\cdot \Delta \mathbf{v} \mid >  \left|\mathrm{Im}\left(\omega(\mathbf{k})\right)\right|$ , the instability is deemed kinetic for most $k$'s, leading to a significantly slower energy loss. This investigation shows that that in addition to plasma heating via instability losses, quasilinear relaxation phase characterised by the momentum diffusion within the beam can alter the initial beam distribution function and energy partition in the beam-plasma system.

At the beginning of the linear growth phase, the spectral energy density in the background plasma grows exponentially through unstable modes. Furthermore, the pair beam undergoes relaxation, which is apparent from the energetic broadening of the beam at a later stage. According to the quasilinear treatment of beam relaxation, diffusion in the momentum space will continue broadening the beam until the beam is completely relaxed, i.e., change in the angular width of the beam approaches the initial beam width $\Delta \theta \sim \theta_{0}$ where $\theta_{0}$ is the initial opening angle of the beam \citep{brejzman1974powerful}. The overall beam evolution is determined using the Fokker-Planck equation, Eq.~\ref{eq:fp}, where the drift term describes the energy drain owing to instabilities and relaxation is governed by the diffusion term. 

Both the analytical solution to Eq.~\ref{eq:fp} for a 1D narrow Gaussian input beam as delineated in Section \ref{subsec:1Danal}, and the numerical solution for a 2D narrow Gaussian input beam as described in Section \ref{subsec:2Dnum} suggest that the energy loss and self-heating of the beam are driven exponentially, respectively with exponents twice and equal to the growth rate in units of plasma frequency. In conjunction with PIC simulations, this investigation indicates that even in the case of a narrow injected beam, where the beam is expected to lose energy initially in the reactive regime, it eventually broadens and the instability growth rate drops. In the final stages of the beam evolution, nonlinear damping effects become important which eventually leads to a saturation of the growth phase, thus stabilizing the beam.

The 2D version of EPOCH resolves space in two dimensions and momentum in three dimensions. Since the instability involves a modulation of the momentum distribution as a function of position, the momentum distribution can only evolve in dimensions that are also resolved spatially. In addition, we performed 3D simulations resolving all three spatial dimensions with a lower resolution, in order to confirm that the qualitative behavior of the beam relaxation is consistent with the above. In both cases, we could show that the change in the momentum distribution (as defined by $\Delta \mathrm{Mean}\left( p_\parallel \right)$, $\mathrm{Std}\left( p_\parallel \right)$, $\mathrm{Std}\left( p_\bot \right)$) occurs at the same time, while the change in the mean momentum grows as twice the growth rate of the electric field, and the momentum widths grow according to the growth rate of the electric field. 

The investigation in the idealized conditions of the PIC simulation assumes a homogeneous cold pair beam propagating through a homogeneous cold plasma. However, both the astrophysical scenario and a laboratory setup can deviate from these simplified assumptions. Inhomogeneity in the plasma acts on the characteristic length scale $L_{\mathrm{inh}}$ which in the direction parallel to beam propagation is given by

\begin{equation}
    L_{\text{inh},\mid \mid} \sim \left| \frac{\partial \mathrm{ln}\,n_\text{p}}{\partial z} \right| ^{-1}.
\end{equation}

In presence of inhomogeneity, relaxation can proceed only if the initial angular width of the beam is smaller than a critical value set by the inhomogeneity length scale, i.e., $\theta_{0} < 1/(\eta_{\mid \mid} \Lambda)$. Here we follow the conventions of \citep{brejzman1974powerful}, where for nonlinear processes $\Lambda = 1$ and in the quasilinear regime $\Lambda = 10$. The parameter $\eta_{\mid \mid}$ is defined as

\begin{equation}
    \eta_{\mid \mid} = \frac{c}{\omega_\text{p} L_{\text{inh},\mid \mid}} \frac{\gamma}{\alpha}.
\end{equation}
However, when the broadening of the beam leads to an increase in the angular spread such that the angular width of the beam $\theta \sim 1/(\eta_{\|} \Lambda)$ the beam stops expanding (\cite{breizman1971quasilinear}, \cite{perry2021role}). If we limit our calculation to the quasilinear case, the above condition for beam relaxation reads
\begin{equation}
    L_{\text{inh},\mid \mid} \gtrsim 1~\text{cm}\,\left(\frac{\Lambda}{10}\right)\left(\frac{10^{-3}}{\alpha}\right)
    \left(\frac{10^{16} \mathrm{cm}^{-3}}{n_\text{p}}\right)^{1/2}
    \left(\frac{\mathrm{Std}\left(p_\bot\right)}{5\,\text{keV}}\right).
\end{equation}
 A corresponding inhomogeneity lengthscale in the transverse direction can be constructed and determined in the same manner.

In addition, a beam created in a laboratory experiment will have finite longitudinal and transverse extensions. Consequently, a beam that is too short or too narrow will not be able to undergo collective effects. A neutral pair beam can be produced in a laboratory by hitting a high-density material such as lead or tungsten with an electron beam, such that electron-positron pairs are produced via a cascade of Bremsstrahlung and pair-production \citep{sarri2015}. This beam will deviate from the idealized picture of a homogeneous beam with a momentum distribution that is independent of position. Particles at the front of the bunch will have higher energies and particles with higher energies will propagate with a smaller opening angle with respect to the beam axis. In addition, the cascading process cannot produce a perfectly neutral beam, since at low energies, in particular, below the critical energy of the material, ionization is an important energy loss mechanism for charged particles producing additional electrons. In this investigation we are interested in a regime of a very dilute beams perturbing a background plasma. If the energy density of the beam approaches the rest mass energy of the background plasma ($\alpha\gamma \approx 1$), this is no longer satisfied. Scenarios such as this can lead to an anomalous gain of energy for the background plasma that can exceed the energy gain of the electromagnetic fields by orders of magnitude, unlike astrophysical pair beams. For laboratory astrophysics experiments, beam parameters such as charge neutrality and bunch length play a significant role in addition to the energy partition of the evolving beam-plasma system.

For the instability process to unfold in a laboratory experiment the beam has to be kept nearly constant for the time the instability grows or equivalently to a distance of propagation. The propagation distance to grow the instability is proportional to the growth rate $\delta$, and is also sensitive to the background plasma density. For parameters $\alpha = 10^{-3}$, $\gamma = 10$ and $n_\mathrm{p} = 10^{16}\,\mathrm{cm}^{-3}$ this distance is roughly $L_{\text{inst},||} \sim$ 1.5 cm. Keeping the beam density constant and increasing the background density leads to a decrease in physical propagation distance, since $\omega_\text{p} \propto n_\mathrm{p}^\frac{1}{2}$ but $\delta \propto \alpha^\frac{1}{3} \propto n_\mathrm{p}^{-\frac{1}{3}}$, such that the physical growth rate $\delta\omega_\text{P}\propto n_\text{p}^\frac{1}{6}$. Furthermore, for a beam of finite length and width, as is the case of a laboratory experiment, increasing the background density would increase the length and width in units of the plasma wavelength, having the beam closely resemble the infinitely extended beam as outlined in this article.

We have established the parameters for a laboratory beam-plasma system and perform PIC simulation for a cold neutral pair beam. In order to seed an oblique instability in the laboratory, a positive slope in the longitudinal momentum distribution of the input pair beam is required. Upon creation of the pair beam by irradiation of a high-$Z$ conversion target with a high energy particle beam, where electron-positron pairs are created via Bremsstrahlung-photons undergoing pair production, the resulting spectrum would feature a large transverse divergence and a broad energy spectrum (first demonstrated in \cite{sarri2015} and further investigated in \cite{sarri2018}). In order to enforce a positive momentum slope and sufficiently improve the transverse beam quality, a combination of beam optics and collimators could be used for beam transport and filtering. The underlying principle is the chromaticity of the magnetic elements, which leads to an energy-dependent transverse focus along the beam axis, similar to  \citep{fuchs2009}. Simulations have shown that a pair beam similar to the cold beam used as input in the PIC simulations can be generated using this filtering mechanism.

We note that while we study the unstable behavior of neutral pair beams, in the absence of external magnetic fields the unstable behavior of a pure electron beam of the same density would be identical up to effects due to the finite current of the electron beam. PIC simulations show that for an infinitely wide electron beam the widening of the momentum distribution is the same for a neutral pair beam as well as a pure electron beam. For a laboratory experiment it could be easier to use an electron beam that is easier to handle as a stand in for the neutral pair beam. 

The presence of unstable behavior in a beam can be established by measuring the broadening of the initially cold momentum distribution which can be easily measured with a magnetic spectrometer if the instability can be grown until its saturation level. The scaling of the saturation level with regards to the beam density ratio $\alpha$ and the beam Lorentz boost $\gamma$ further establish that the effect on the momentum distribution is caused by the instability. 

Extrapolating our results to density contrasts and Lorentz factors relevant for astrophysical scenarios suggests that in astrophysical beams the total energy loss of the beam before saturation becomes negligibly small and can thus not significantly alter the secondary TeV gamma ray flux. The momentum diffusion can lead to an angular broadening of the beam not unlike the effect of deflections in a large-scale magnetic field. However, extrapolating Eq.~\ref{eq:theta_sat} suggests that this broadening is significantly smaller than expected magnetic deflection and experimental angular resolutions. 

In the astrophysical scenario, however, fresh electrons and positrons are continuously injected into the beam via pair production. Such pairs have not yet been affected by the momentum broadening and are thus colder in the perpendicular direction. This colder beam component could potentially further drive the instability. As a consequence, the linear instability growth could continue for a longer timescale before the onset of the saturation, compared to that derived from our PIC simulations. 

\section{Summary and Outlook} \label{sec:conclusion}

In this article, we outline the evolution of relativistic dilute neutral pair beams consisting of electrons and positrons propagating through a cold plasma through collective plasma processes. In the initial phase, the beam undergoes Langmuir oscillations that embody the exponential growth of the spectral energy density of the plasma through instabilities. In the absence of magnetic field, the dominant instability is the oblique electrostatic instability and its growth rate derived using linear perturbation theory holds true. In the following phases, the instability drives the beam into a state of diffusive relaxation through electric field fluctuations that scatters the beam pairs. The growth of modes continue but it takes place in the kinetic regime even for pair beams that are near-monochromatic at injection. We demonstrate that the instability losses and beam relaxation can be respectively represented by the exponentially growing drift and diffusion terms of a compact 2D Fokker-Planck prescription. While the predictions of the linear theory to describe the growth of the unstable modes and the quasilinear treatment of the energetic widening of the pair beam are limited up until the onset of nonlinear saturation of instabilities, the widening of the momentum distributions can be shown to proceed exponentially until a saturation is reached. In addition, results from the PIC simulations show that relaxation can continue with a $\sqrt{t}$ scaling after the saturation of growth of the field energy density, leading to an increase in the beam opening angle.

We point out major differences between the laboratory and astrophysical scenarios and extrapolate the results of the PIC simulation to describe the fate of the blazar-induced pair beams. Inhomogeneities in the plasma can suppress the  growth of instabilities and stall beam relaxation. Initial correlation between the parallel and perpendicular momentum in the beam distribution function can also significantly alter the evolution of the beam. Another important consideration would be nonlinear damping effects that lead to a saturation of growth for the unstable modes. In order to better understand the modification of the pair beam spectrum owing to collective plasma effects, in a real-life laboratory setup, finite beam extension in the longitudinal and transverse direction, extent of charge neutrality violation, chirp and beam emittance could be measured. It is also worthwhile to explore the evolution of pair beams in magnetized plasma in order to gain further insights into the TeV pair beams propagating through the IGM. Since the exploration of the above is beyond the scope of our current work, we defer them to future publications.

%\appendix

\section*{Acknowledgement}

\begin{acknowledgments}
\begin{nolinenumbers}
MB, OG, RDS, and BZ were funded by the Deutsche Forschungsgemeinschaft (DFG, German Research Foundation) under Germany’s Excellence Strategy – EXC 2121 ``Quantum Universe'' – 390833306. OG is supported by the European Research Council under Grant No. 742104 and by the Swedish Research Council (VR) under the grants 2018-03641 and 2019-02337.  CBS was supported by the Office of Science, Office of High Energy Physics, of the U.S. Department of Energy under Contract No. DE-AC02-05CH11231. We thank Martin Lemoine for useful comments on the manuscript.
\end{nolinenumbers}
%This research was supported in part through the Maxwell computational resources operated at Deutsches Elektronen-Synchrotron DESY, Hamburg, Germany. 
\end{acknowledgments}

\bibliography{main}{}
\bibliographystyle{aasjournal}



\end{document}